\documentclass{article}
\usepackage{stmaryrd}
\usepackage[english]{babel}
\usepackage{framed}
\usepackage[letterpaper,top=1in,bottom=1in,left=1in,right=1in]{geometry}
\usepackage{xypic}
\usepackage{mdframed}
\usepackage{amsmath}
\usepackage{graphicx}
\usepackage{amssymb}
\usepackage{mathpartir}
\usepackage{irule}
\usepackage{tikz}
\usepackage[colorlinks=true, allcolors=blue]{hyperref}
\usepackage{tikz-cd}
\def\Inst{{\bf Inst}}
\def\iinst{{{ \textendash}\bf \Inst}}
\def\taking{:}
\newcommand{\To}[1]{\xrightarrow{#1}}

\newcommand{\pullbackcorner}[1][dl]{\save*!/#1-1pc/#1:(-1,1)@^{|-}\restore}

\title{Consensus-Free Spreadsheet Integration}
\author{Brandon Baylor, James Hansen, Esteban Montero \\ Chevron \and Eric Daimler, Ryan Wisnesky \\ Conexus AI}

\begin{document}
\maketitle

\begin{abstract}
We describe a method for merging multiple spreadsheets into one sheet, and/or exchanging data among the sheets, by expressing each sheet's formulae as an algebraic (equational) theory and each sheet's values as a model of its theory, expressing the overlap between the sheets as theory and model morphisms, and then performing ``colimit'',  ``lifting'', and ``Kan-extension'' constructions from category theory to compute a canonically ``universal'' integrated theory and model, which can then be expressed as a spreadsheet.  Our motivation is to find methods of merging engineering models that do not require consensus (agreement) among the authors of the models being merged, a condition fulfilled by our method because theory and model morphisms can be mechanically checked to be semantics-preserving.  We describe a case study of this methodology on a real-world oil and gas calculation at a major energy company, describing the theories and models that arise when integrating two different casing pressure test (MASP) calculation spreadsheets constructed by two non-interacting engineers.  We also describe the automated theorem proving burden associated with both verifying the semantics preservation of the overlap mappings as well as verifying the ``conservativity''/``consistency'' of the resulting integrated sheet.  We conclude with thoughts on how to apply the methodology to scale engineering efforts across the enterprise.
\end{abstract}

\section{Introduction}

As the energy industry moves to model-based software for performing day-to-day engineering tasks, it becomes more and more important to ensure the semantic consistency of models composed of related models (integrated models).  For example, we don't want errors to propagate from one model to another or to try to integrate models with conflicting requirements (e.g. requiring positive and negative voltages on the same wire at the same time).  Commonly, to ensure semantic consistency of integrated models the human subject matter experts that created them communicate informally with each other about their respective understandings of the integrated model and check for consistency of the integrated model with respect to their original models (think NASA mission control with its various `functions').  Although using groups of human experts to certify/construct integrated models works for small numbers of input models, this methodology is both costly and not scalable, because in principle, all the humans may need to communicate with each other (not everything can pass through mission control), and moreover, people may disagree about the meaning of the input and/or integrated models.  Therefore, to construct/certify engineering models in a scalable way, we must make sure that when models are merged, the original authors of the models need not be involved.  

\subsection{Contributions}

In this paper, we propose a methodology that does precisely the above for many spreadsheet-based engineering models, based on treating each sheet as both an algebraic (equational) theory~\cite{10.5555/280474} and a model of that theory (a so-called ``olog''~\cite{10.1371/journal.pone.0024274}) and using techniques from automated theorem proving~\cite{10.5555/280474} and category theory~\cite{MR2229319} (summarized in~\cite{schultz_wisnesky_2017}) to construct composed ologs (theories and models), which can then be exported as spreadsheets.  Relationships between input ologs are captured as theory and model ``morphisms''~\cite{conf/wadt/MossakowskiKM14} (olog morphisms), for which we generate and solve~\cite{schultz_wisnesky_2017} verification conditions to ensure they are semantics-preserving without recourse to the original spreadsheet authors.  We demonstrate the practicality of our methodology on a maximum anticipated surface pressure (MASP) calculation done in two different spreadsheets by two non-interacting engineers, describing the associated ologs as well as the verification conditions generated by both relating the models and integrating them.  Specifically, our research contributions (new results) are:  
\begin{itemize}
    \item a method to represent some spreadsheets as algebraic theories~\cite{10.5555/280474} and models (`ologs'~\cite{10.1371/journal.pone.0024274}), and vice versa; and, 
    \item a case study about two independently-developed ologs corresponding to a standard spreadsheet-based MASP calculation, as well as another olog and two olog morphisms~\cite{conf/wadt/MossakowskiKM14} capturing the relationship of the first two ologs to each other; and,
    \item a use case about how to apply the technology demonstrated in the case study to reducing costs in auditing spreadsheet-based models for requirements compliance.
    
\end{itemize}

\subsection{Outline}
In the remainder of the introduction, we provide background on MASP calculations, their applications, and why integrating/exchanging data between them can be useful.  The rest of this paper is structured as follows:

\begin{enumerate}
\item In section~\ref{sec:ologs} we describe how spreadsheets can be represented as ologs and vice-versa.
\item In section~\ref{sec:integration} we describe the two source sheets/theories/models of MASP calculations to be merged, the overlap between them, and integrated result and its applications.  We also discuss the practicalities of the case study.
\item In section~\ref{sec:vc} we describe the verification conditions generated by relating the sources to each other and during construction of the integrated result.  This section will be of interest primarily to computer scientists. 
\item In section~\ref{sec:conc} we conclude by discussing applications of our methodology across the enterprise.  

\end{enumerate}

It is our hope that our methodology can be applied by engineers without any training in category theory, logic, or algebra, and that this paper should be readable without such knowledge as well.  However, to fully understand this paper, knowledge of all three areas is required, and to that end, we have included appendices that review them (Appendices \ref{sec:cat}, \ref{sec:alg}, and \ref{sec:cql}, respectively).  

\subsection{MASP Background}

Maximum anticipated surface pressure (MASP) is the highest (worst case) surface pressure that is expected to be encountered throughout the well construction process \cite{IADC}.  MASP calculation is a key engineering activity for both preliminary and detailed well design because it impacts well control equipment specifications and pressure test plans. In particular, MASP determines the sizing and ratings for pressure control/containment equipment such as blowout preventers and strings of casing \cite{CFR2021}. The results of the MASP calculation are also used to derive numerical values to test said pressure control/containment devices in the field. These are key aspects of maintaining process safety and integrity of oil and gas wells during drilling, completion, or intervention operations.

MASP is normally calculated for each casing section as part of the well design process. This typically assumes a shut-in well and various types of fluid in the wellbore (e.g., mud or gas). Guidance for determining the appropriate calculation scenarios (e.g., the correct ratio of fluid in the well) are often provided by global standards and technical requirements. Other inputs include pore pressure and fracture gradient data (obtained from offset wells or regional analogues), fluid density values, well depths, and casing specifications. 

\begin{figure}
\includegraphics[width=6in]{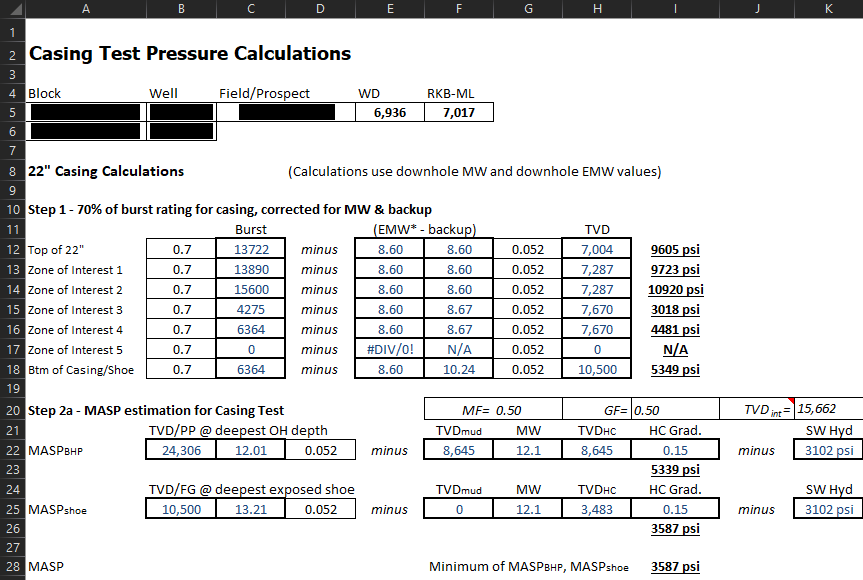}

\caption{Original MASP Excel Sheet}
\label{fig:originalMASP}
\end{figure}

The determination of MASP typically includes pore pressure uncertainties, new geologic zones of interest being added/encountered, regulatory requirements, and other unanticipated factors. For these reasons there are verification activities to assure/audit that engineering calculations are sound and compliant with technical requirements. Moreover, there are often several engineers working on singular aspects of the well plan and then collaborating to ensure an optimal design. In either case, auditors or engineers may want to integrate their respective viewpoints to generate new insights and ensure explicit alignment between each other, suggesting use cases described in the conclusion. 

\section{Ologs and Spreadsheets}
\label{sec:ologs}

The primary research contribution of this paper, besides the case study, is a methodology for representing certain spreadsheets as ologs~\cite{10.1371/journal.pone.0024274} (theories and models), which we describe in this section.   We begin by observing that it is trivial to represent every spreadsheet as a model~\cite{schultz_wisnesky_2017}  of a signature containing one constant symbol for every used cell.  For example, the initial model of the theory:
\begin{verbatim}
b17 c12 : Float    b17 = .07    c12 = 13722
\end{verbatim}
formalizes a fragment of the MASP sheet in Figure~\ref{fig:originalMASP}.  In contrast, this paper's method is based on leveraging the typical, natural decomposition of most sheets into smaller tables into order to provide additional semantics (sorts and symbols and equations).  First, we define a notion of {\it categorical normal form} for a sheet, from~\cite{DBLP:journals/iandc/Spivak12}. Then we define translations to/from olog sheets in this normal form.  A spreadsheet is in this form when:

\begin{itemize}
    \item the sheet is composed entirely of rectangular sub-tables, each with a name, a  set of named columns, and a distinguished ``primary key'' column. The cells in the primary key column are called the row-ids of t and semantically row-ids are meaningless identifiers, not data;
    \item for every column $c$ of a table $t$ there is either some table $t'$ such that the value in each cell of $c$ refers to some row-id of $t'$ (in which case we say that $c$ is a ``foreign key'' column from $t$ to $t'$; or column $c$ of table $t$ is a ``pure data'' / ``value'' column of $t$ with values in some non-row-id type such as Float.  
\end{itemize}

Of course, most spreadsheets are not in categorical normal form to begin with.  Fortunately, we have found that human subject matter experts (not computer scientists) can rapidly and accurately create new spreadsheet tabs that are in the above form and that reference back into the original sheets.  These new tabs provide an auditable ``olog view'' of the original sheet, and heuristic methods to automatically generate these olog tabs are a subject of current work.  
In our case study, two engineers (paper co-authors Brandon and James) both independently created olog tabs from the government-provided MASP calculation spreadsheet shown in Figure~\ref{fig:originalMASP}.  In practice, each engineer may start with a different spreadsheet, but the olog views of the two engineers for this one government sheet are so different that for the purposes of this paper there is no difference between starting from one sheet or two: what matters is starting from two different ologs.     

{\bf Remark.} In this paper we will always be working with spreadsheets that correspond to models in the sense of logic~\cite{schultz_wisnesky_2017}, but the CQL software that we are using~\cite{cql} allows us to work with ``presentations''  of spreadsheets (those that are missing values in foreign key columns; missing values in data columns don't matter here) by automatically constructing initial models of equational theories for us.

\subsection{Spreadsheet to Olog}

The above categorical normal form suggests an easy way to generate a signature (sorts and symbols) from a spreadsheet, especially when column headers are used to indicate the foreign keys directly in Excel.  However, categorical normal form doesn't suggest any axioms along with the generated signature.  The way in which we recover axioms is by examining the formulae of the sheet.  In particular, to be imported as a universally quantified equation, we require that a column's rows all contain ``the same'' formula-- a formula that is a function of only its row -- which itself must be written entirely of spreadsheet functions applied to ``lookups'' of foreign-key columns.  For example, in the MASP olog from Source A, the MASP Calc. Step 1 table contains a column called 70\% Burst (not-corrected) (column C) whose rows have the form: 

\begin{verbatim}
C76 = LOOKUP(E76, $A$13:$A$41,$F$13:$F$41)*B76 
C77 = LOOKUP(E77, $A$13:$A$41,$F$13:$F$41)*B77
...
\end{verbatim}
where column E has ``Casing Section'' as a header and column B has ``De-rated Percent'' as a header.  Moreover, the indicated ranges are those of the ID and ``Burst Rating'' column in another table.  As such we import the following equation (also seen in Figure~\ref{fig:eqsA}): 
\begin{verbatim}
forall x:"MASP Calc. Step 1", "70% Burst (not corrected)"(x) =
		    "Burst Rating"("Casing Section"(x)) * "De-Rated Percent"(x)
\end{verbatim}

In the above example our columns all produce numbers.  In general, however, columns may also produce boolean values indicating the equivalence of two arbitrary expressions.  We say that such columns are not ``definitional'' in the sense of the 70\% burst column in the example above.  The input MASP calculations for this paper contain entirely ``definitional'' formulae, but as described later the integrated result contains some columns that are not definitional (the integrated sheet contains some boolean-valued columns that constrain other columns).  

\subsection{Olog to Spreadsheet}

With one caveat, to convert from ologs to spreadsheets it is necessary to merely invert the representation in the above section.  The caveat is that the above discussion does not describe how to encode ``type algebras''~\cite{schultz_wisnesky_2017} -- missing values and equations between missing values that can appear in ologs but not in spreadsheets that people construct naturally.  To encode a non-trivial type algebra, we must include a single-column table in the spreadsheet for each type, such as integer and string.  Then, each ``skolem variable''/missing value at that type becomes a row in this new table, and formulae may refer to this new cell (such formulae will compute as N/A in Excel, of course).  Finally, for each type we also require another table with two columns wherein ground equations between these missing values are encoded as in the preceding section.  An example is shown below. 

\begin{verbatim}
  A        B    C   D        E  F
1 Person   Age      Integer     IntegerEqs 
2 p1       20       x           =(D2+D3=20)
3 p2       =D2      y
4 p3       =D3
\end{verbatim}

Of course, when we import a spreadsheet containing a type algebra as encoded above, we must be careful to import the type tables as a type algebra, and not as ``user tables''. In this paper's example, the integrated result does have a non-trivial type algebra, because it contains  blank cells in calculated columns in the original sheets. Whether our result's type algebra is contradictory is discussed in Section~\ref{sec:vc}.




{\bf Remark.}  Our olog to spreadsheet and reverse translations fail on empty ologs because there are no corresponding spreadsheet cells with which to record equations.  This problem can be remedied by including, for each empty table, a ``phantom'' row that records the equations with non-formulae cells left blank.  Or, by writing the equations down as additional column headers.  

\subsection{Free Theories Not Closed Under Colimits}

We might ask, ``if we start with two definitional/free sheets, turn them into ologs, compose them using colimits as described later, and then turn the result back into a sheet, will the resulting sheet be entirely definitional/free as well''? That is, if we start with sheets where every column is or computes a number will we end up with a sheet where every column is or computes a number? The answer is no: in mathematical terms, a free theory such as $f(x)=x^2$ and a free theory such as $g(y)=y^3$ can be merged along the free theory $f(z) = g(z)$ to give the theory in which $x^2=x^3$, which is non-free (non-definitional).  A non-free theory requires boolean-valued columns to represent as a spreadsheet and requires more effort to reason about than a free theory but is otherwise benign\footnote{For example, spreadsheets can only compute initial models of free/definitional theories, implying that there is no spreadsheet that can compute the algebraic merge of two other spreadsheets-- the merge must be done externally to Excel (where it can, for example, create an infinite number of rows or fail with contradictions) and then re-expressed as a sheet.}.  In our MASP example, failure of freeness preservation under colimits  manifests when we have multiple equations that define the same symbol, such as ``Well'' below: 
\begin{verbatim}
forall x,     x.Well = x."TVD Shoe".Well
forall x,     x.Well = x."TVD Deepest OH".Well
\end{verbatim}
The theory above is not free because it contains non-trivial equations such as 
\begin{verbatim}
forall x,   x."TVD Shoe".Well = x."TVD Deepest OH".Well
\end{verbatim}
and so when exported as a spreadsheet will contain a boolean-valued column witnessing the truth of the above equation; this boolean-valued column is considered part of the schema, not part of the data, of the sheet's corresponding olog.  For example, any sheet containing the equations above would be exported as
\begin{small}
\begin{verbatim}
    Step1  Well  TVD Shoe   TVD Deepest OH  x.Well = x."TVD Shoe".Well  x.Well = 
                                                                          x."TVD Deepest OH".Well    
1   s1     w1     x1         y1             true                        true
2   s2     w2     x2         t2             true                        true
3   ...   
\end{verbatim}
\end{small}
where each \texttt{true} is obtained by evaluating the formula in the column it appears in, where that formula will contain lookups etc as described earlier.

A related question is whether the merge of two finite sheets is always finite, where the answer is also no, such as when the schema $\cdot \to_p \cdot$ is merged with the schema $\cdot \leftarrow_q \cdot$ along the left and right dots, resulting in a schema with a cycle that is not ``broken'' by equations $p(q(x))=x$ and $q(p(x))=x$ and thus which will force the integrated sheet to be infinite when there is no data level overlap of the sources (having e.g. a row $c$, and a row $c.p$, and a row $c.p.q$ not equal to $c$, and so on).  In such a case it is customary to add additional equations to break the cycle and force a finite model.

\subsection{Axiomatizing Spreadsheet Functions}

In this subsection we axiomatize a small collection of spreadsheet functions that are sufficient for our case study; the resulting theory is the {\sf Type} input in the algebraic integration design pattern in Figure~\ref{fig:algint}. We start with an infinite ground (variable-free) equational theory for addition {\tt +}, subtraction {\tt -}, multiplication {\tt *}, maximum {\tt MAX}, and minimum {\tt MIN}, all of which are binary operations on floating point numbers:
$$
0{\tt +}0 = 0 \ \ \ \ 0{\tt +}1=1 \ \ \ \ 1{\tt +}2.2 = 3.2 \ \ \ \ \ldots
\ \ \ \ 
0 {\tt *} 0 = 0 \ \ \ \ 0 {\tt *} 1 = 0 \ \ \ \ 1 {\tt *} 2.2 = 2.2 \ \ \ \ \ldots
\ \ \ \ 
{\tt MAX}(3.5,2)=3.5
\ 
\ldots
$$

From there, we add the non-ground (universally quantified) axioms stating that $(0,1,+,\times)$ form a commutative ring~\cite{MR2229319}, as well as some axioms stating basic properties of how {\tt MAX} and {\tt MIN} interact with arithmetic.  




{\bf Remark.} In this paper, we have associated a type with each spreadsheet cell, either floating point or string.  However, in many spreadsheets, cells are not typed, and functions such as addition will throw errors on non-numeric inputs.  This dynamically-typed situation can also be represented in the manner of this section, with a ``universal type''.  Multiple types also allow for the possibility of representing different units (feet, inches, etc) as different types, reducing risk of error.

{\bf Remark.} One of the original sheets used division in a single column, which for expediency we removed in favor of treating the column as plain data so as not to need to deal with division at all in this case study.
 


\section{MASP Sheet Formalizations and Integration}
\label{sec:integration}

We now recall the algebraic data integration design pattern~\cite{schultz_wisnesky_2017}, displayed in Figure~\ref{fig:algint}.  Our goal in this section is to describe all of the inputs and some of the outputs in the above diagram, including how they were constructed, either by human or machine, for our MASP example.  

{\bf Remark.} Intuitively, the guarantee provided by algebraic data integration is that the output of our methodology ``is at least as good as every solution'' to the problem of integrating the input ologs and overlap.  Universality also guarantees a complete ``lineage'' or ``provenance'' of how every output row was constructed, although we do not elaborate further in this paper.

\begin{figure}
    \begin{minipage}[t]{0.3\textwidth}
\vspace{-.35in}
\[
\begin{tikzcd}[column sep=small]
S_O\ar[rr, "M_A"]\ar[dd, "M_B"']\ar[dr, "I_O" description, "" name=OR, ""' name=OL]&&
S_A\ar[dd, "N_A"]\ar[dl, "I_A" description, "" name=AR, ""' name=AL]\\&
\mathsf{Type}\\
S_B\ar[rr, "N_B"']\ar[ur, "I_B" description, ""' name=BR, "" name=BL]&&
S\ar[ul, "I" description, "" name=R, ""' name=L]
\ar[from=OL, to=OL|-BL, shorten <=3pt, shorten >=3pt, Rightarrow, "h_B"']
\ar[from=OR, to=AL|-OR, shorten <=3pt, shorten >=3pt, Rightarrow, "h_A"]
\ar[from=AR, to=L-|AR, shorten <=3pt, shorten >=3pt, Rightarrow, "k_A"]
\ar[from=BR, to=BR-|R, shorten <=3pt, shorten >=3pt, Rightarrow, "k_B"']
\end{tikzcd}
\]
Universality guarantees that for any other $S'$, $N'_A, N'_B$, $I'$, $k'_A,k'_B$ making the above diagram commute, there is a unique schema mapping $M:S\to S'$ and data mapping $k:I\to I'$ such that $N'_A = N_A; M$ and $N'_B = N_B; M$ and $k'_A = k_A ; k$ and $k'_B = k_B ; k$. 
\end{minipage}
\begin{minipage}[t]{0.7\textwidth}
\begin{itemize}
    \item {\sf Type} is an algebraic (equational) theory representing types and functions (addition, subtraction, etc.).
    \item $S_O$, $S_A$, $S_B$ are input schemas/theories (with $S_O$ the overlap) and $S$ is the output integrated schema/theory (excel formulae).
    \item $I_O : S_O \to {\sf Type}$, $I_A : S_A \to {\sf Type}$, and $I_B : S_B \to {\sf Type}$ are input models/databases (with $I_O$ the overlap) and $I : S \to {\sf Type}$ the output integrated model/database (excel values).
    \item $M_A : S_O \to S_A$ and $M_B : S_O \to S_B$ are input schema mappings/theory morphisms, and $N_A : S_A \to S$ and $N_B : S_B \to S$ are output schema mappings/theory morphisms (functors).
    \item $h_A : I_0 \Rightarrow M_A;I_A$ and $h_B : I_O \Rightarrow M_B;I_B$ are input data mappings/modem morphisms, and $k_A : I_B \Rightarrow N_A;I$ and $ k_B : N_B;I$ are output data mappings/model morphisms (natural transformations). \\
\end{itemize}
\end{minipage}
    \caption{Algebraic Schema and Data Integration}
    \label{fig:algint}
\end{figure}

\subsection{Schema Integration}

In this section, we assume that the input spreadsheets have been converted to ologs as described in the previous sub-sections and discuss the resulting schema (theory part) of the olog conversion.  Recall that schema integration is the process of taking input schemas (theories) $S_O$, $S_A$, $S_B$ and computing output schema (theory) $S$, according to the usual algebraic integration diagram in Figure~\ref{fig:algint}.  

\subsubsection{Schema A}

\begin{figure}
\includegraphics[width=6.5in]{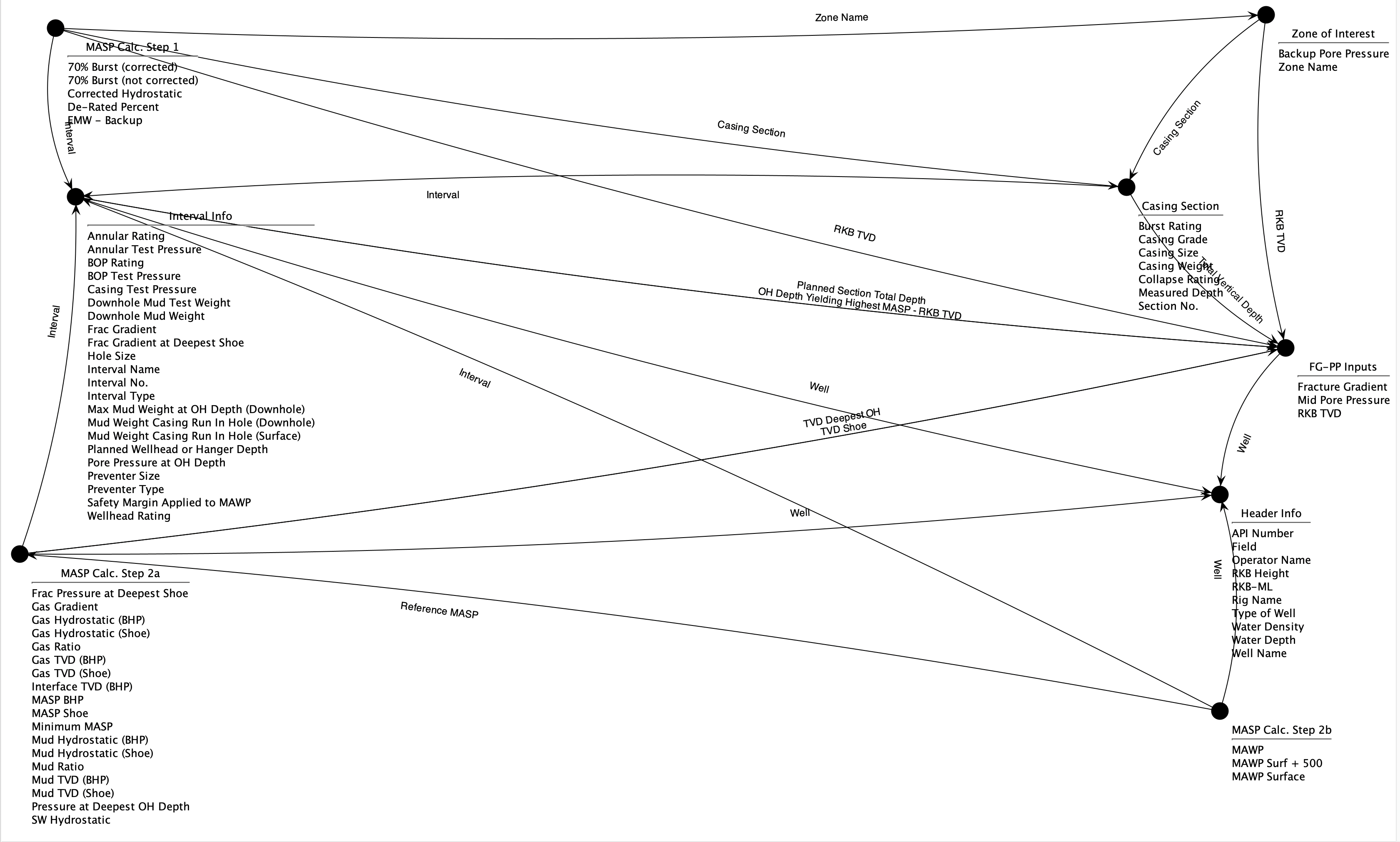}

\caption{Signature for Schema A}
\label{fig:schemaA}
\end{figure}

The signature (sorts and symbols) of schema A are defined in Figure~\ref{fig:schemaA}.  Each edge denotes a unary function between tables and each row in the table associated to a node is a unary function targeting a type (String, Int, etc) (i.e., a ``data column'').  The equations for Schema A are defined in Figure~\ref{fig:eqsA}.

Schema A contains information about MASP as defined by the Header Info table in Figure~\ref{fig:schemaA}. This sets a broad context for how MASP is being calculated along with high level data that applies to all hole sections where the engineering analysis is performed. These specific hole sections are defined by multiple tables: ``Casing Section'', ``Zone of Interest'', and ``Interval Info''.  These tables reflect the general well construction process and ensure alignment between equipment specifications (such as casing) and their associated depths (where formation pressure inputs apply). The specific pressure inputs, correlated by  well depth, are defined in the FG-PP Inputs table and are usually provided by teams working in the subsurface domain. Finally, the sequence of the MASP calculation itself is defined by the following tables: ``MASP Calc. Step 1'', ``MASP Calc. Step 2a'', and ``MASP Calc. Step 2b''. In addition to the computations, these tables include attributes for the appropriate calculation scenarios (such as the ratio of mud or gas in the well).

\begin{figure}    
\begin{footnotesize}

\begin{verbatim}
forall x:"Header Info", "RKB-ML"(x) = "Water Depth"(x) + "RKB Height"(x)

forall x:"MASP Calc. Step 1", "70% Burst (corrected)"(x) =
		     "70% Burst (not corrected)"(x) - "Corrected Hydrostatic"(x)

forall x:"MASP Calc. Step 1", "70% Burst (not corrected)"(x) =
		    "Burst Rating"("Casing Section"(x)) * "De-Rated Percent"(x)

forall x:"MASP Calc. Step 1", "EMW - Backup"(x) =
		    "Downhole Mud Weight"("Interval"(x)) - "Backup Pore Pressure"("Zone Name"(x))

forall x:"MASP Calc. Step 1", "Corrected Hydrostatic"(x) = .052 * ("RKB TVD"("RKB TVD"(x)) * "EMW - Backup"(x))

forall x:"MASP Calc. Step 2a", "MASP BHP"(x) = (("Pressure at Deepest OH Depth"(x) - "SW Hydrostatic"(x))
        - "Gas Hydrostatic (BHP)"(x)) - "Mud Hydrostatic (BHP)"(x)

forall x:"MASP Calc. Step 2a", "MASP Shoe"(x) = (("Frac Pressure at Deepest Shoe"(x) - "Mud Hydrostatic (Shoe)"(x))
        - "Gas Hydrostatic (Shoe)"(x)) - "SW Hydrostatic"(x)

forall x:"MASP Calc. Step 2a", "Mud Hydrostatic (BHP)"(x) =
        .052 * "Mud TVD (BHP)"(x) * "Max Mud Weight at OH Depth (Downhole)"(Interval(x)))

forall x:"MASP Calc. Step 2a", "Gas Hydrostatic (BHP)"(x) = "Gas Gradient"(x) * "Gas TVD (BHP)"(x)

forall x:"MASP Calc. Step 2a", "SW Hydrostatic"(x) = 0.052 * "Water Depth"("Well"(x)) * "Water Density"("Well"(x))
		
forall x:"MASP Calc. Step 2a", "Pressure at Deepest OH Depth"(x) =
        "RKB TVD"("TVD Deepest OH"(x)) * .052 * "Pore Pressure at OH Depth"("Interval"(x))

forall x:"MASP Calc. Step 2a", "Interface TVD (BHP)"(x) =
        "RKB-ML"("Well"(x)) + ("Gas Ratio"(x) * ("RKB TVD"("TVD Deepest OH"(x)) - "RKB-ML"("Well"(x))))
		
forall x:"MASP Calc. Step 2a", "Mud TVD (BHP)"(x) =
        ("RKB TVD"("TVD Deepest OH"(x)) - "RKB-ML"("Well"(x))) * "Mud Ratio"(x)
		
forall x:"MASP Calc. Step 2a", "Gas TVD (BHP)"(x) = 
        ("RKB TVD"("TVD Deepest OH"(x)) - "RKB-ML"("Well"(x))) * "Gas Ratio"(x)
		
forall x:"MASP Calc. Step 2a", "Frac Pressure at Deepest Shoe"(x) =
        "RKB TVD"("TVD Shoe"(x)) * .052 * "Frac Gradient at Deepest Shoe"("Interval"(x))

forall x:"MASP Calc. Step 2a", "Mud TVD (Shoe)"(x) = 
        MAX(0,("RKB TVD"("TVD Shoe"(x)) - "Gas TVD (Shoe)"(x)) - "RKB-ML"("Well"(x)))
		
forall x:"MASP Calc. Step 2a", "Gas TVD (Shoe)"(x) = 
        MIN("RKB TVD"("TVD Shoe"(x)) - "RKB-ML"("Well"(x)), "Interface TVD (BHP)"(x) - "RKB-ML"("Well"(x)))
		
forall x:"MASP Calc. Step 2a", "Minimum MASP"(x) = MIN("MASP BHP"(x),"MASP Shoe"(x))

forall x:"MASP Calc. Step 2b", "MAWP"(x) = 
        "Minimum MASP"("Reference MASP"(x)) + "SW Hydrostatic"("Reference MASP"(x))
		
forall x:"MASP Calc. Step 2b", "MAWP Surface"(x) = 
        "MAWP"(x) - ((.052 * "RKB-ML"("Well"(x))) * "Downhole Mud Weight"("Interval"(x)))
		
forall x:"MASP Calc. Step 2b", "MAWP Surf + 500"(x) = "MAWP Surface"(x) + 500   
    \end{verbatim}
    \end{footnotesize}
    \caption{Equations for Schema A}
    \label{fig:eqsA}
\end{figure}

    \subsubsection{Schema B}
    
\begin{figure}
\includegraphics[width=6.5in]{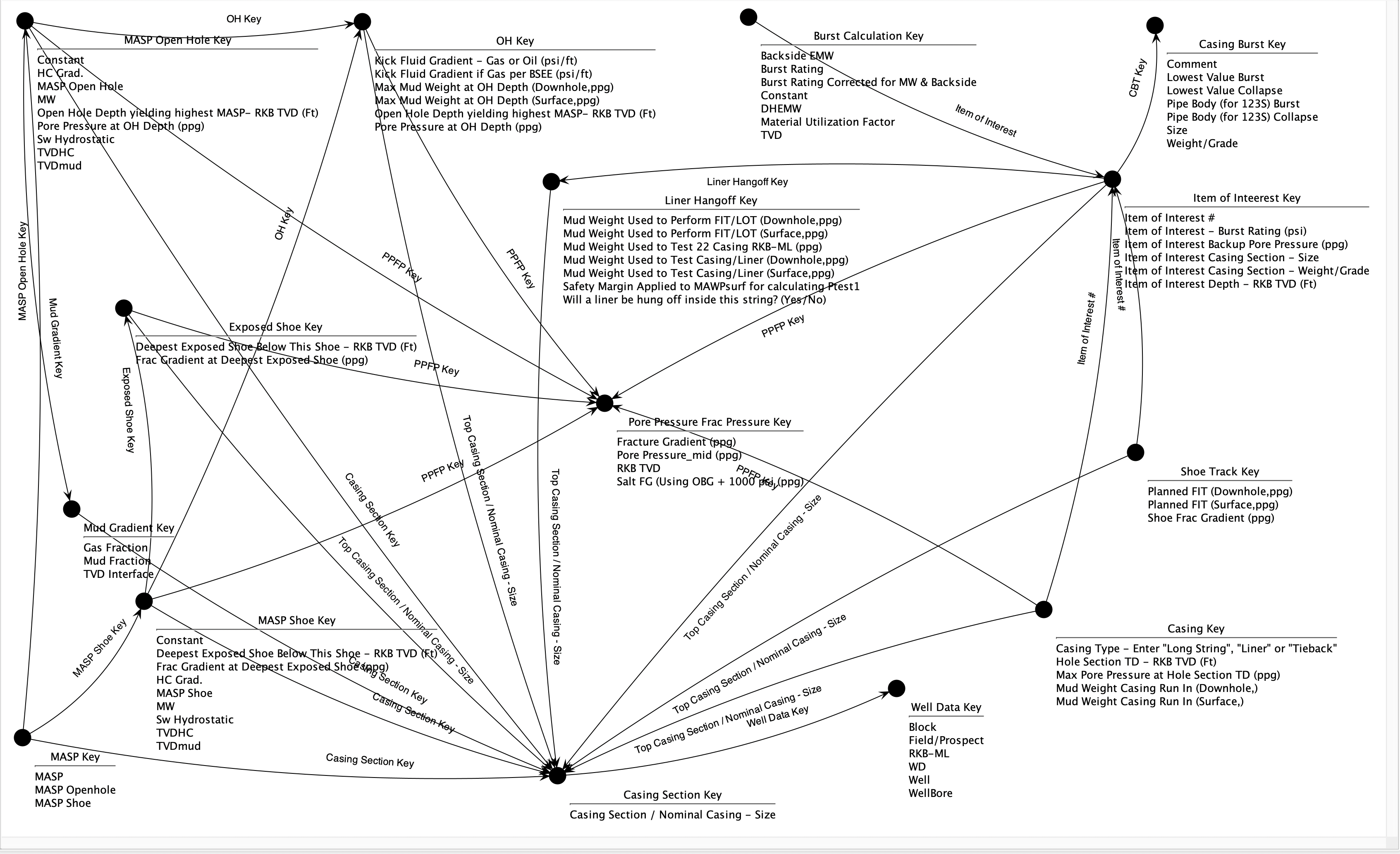}
\caption{Signature for Schema B}
\label{fig:schemaB}
\end{figure}

The signature (sorts and symbols) of schema B are defined in Figure~\ref{fig:schemaB}.  The equations for schema B are defined in Figure~\ref{fig:eqsB}.  

Schema B follows a similar pattern as schema A because they share the same MASP sheet as a starting point. However, because the input models were independently developed there are key differences. Much of the disparity is driven by how the engineers chose to approach the original MASP calculation sheet. Schema A models five hole sections and began olog conversion from the Header Info table. Schema B, on the other hand, approached the olog conversion from the bottom-up and modeled only one hole section starting from the MASP computation itself.

Schema B defines the high level well information in the ``Well Data Key'' table. The particular depths of interest, where the MASP calculation was performed for the lone hole section, are defined by the ``Item of Intereest Key'' table\footnote{Our data sources contain typos in their original schemas and data which are reflected in our development.}. Schema B also captures a series of tables to differentiate information related to the various steps of the MASP calculation: ``Exposed Shoe Key'', ``OH Key'', and ``Casing Burst Key''. The subsurface depth and pressure inputs are defined in the ``Pore Pressure Frac Pressure Key'' table, and the mud/gas scenarios are defined in the ``Mud Gradient Key'' table. The MASP computation is decomposed across the following tables: ``Burst Calculation Key'', ``MASP Open Hole Key'', ``MASP Shoe Key'', and ``MASP Key''. Schema B also includes the ``Exposed Shoe Key'' table, which informs the fracture gradient variables in the MASP shoe calculation. The different MASP uses (open hole and shoe) are separated in Schema B for transparency. Finally, schema B contains tables that contain a few data points that serve more as an efficient modeling structure to map keys between tables (``Casing Section Key'' and ``Casing Key''); this approach was taken to ensure each critical ``variable'' received its own table to better inform future auditing of results.

    \begin{figure}    
\begin{footnotesize}
    \begin{verbatim}
forall x:"Casing Key",  x."Max Pore Pressure at Hole Section TD (ppg)" = x."PPFP Key"."Pore Pressure_mid (ppg)"
		
forall x:"Item of Inteerest Key", x."Item of Interest Backup Pore Pressure (ppg)" = x."PPFP Key"."Pore Pressure_mid (ppg)"

forall x:"OH Key",  x."Pore Pressure at OH Depth (ppg)" = x."PPFP Key"."Pore Pressure_mid (ppg)"

forall x:"Burst Calculation Key", x."Burst Rating" = x."Item of Interest"."Item of Interest - Burst Rating (psi)"
forall x:"Burst Calculation Key", x."Backside EMW" = x."Item of Interest"."Item of Interest Backup Pore Pressure (ppg)"
forall x:"Burst Calculation Key", x."TVD" = x."Item of Interest"."Item of Interest Depth - RKB TVD (Ft)"
forall x:"Burst Calculation Key", x."Burst Rating Corrected for MW & Backside" =
    (x."Material Utilization Factor" * x."Burst Rating") - (x.DHEMW - x."Backside EMW" * x.Constant * x.TVD)

forall x:"MASP Open Hole Key", x."Pore Pressure at OH Depth (ppg)" = x."OH Key"."Pore Pressure at OH Depth (ppg)" 	
forall x:"MASP Open Hole Key", x."MW" = x."OH Key"."Max Mud Weight at OH Depth (Downhole,ppg)" 
forall x:"MASP Open Hole Key", x."HC Grad." = x."OH Key"."Kick Fluid Gradient if Gas per BSEE (psi/ft)" 

forall x:"MASP Open Hole Key", x."MASP Open Hole" =
        -(-(-((x."Open Hole Depth yielding highest MASP- RKB TVD (Ft)" 
        * x."Pore Pressure at OH Depth (ppg)" * x.Constant),
        x.TVDmud * x.MW * x.Constant), x.TVDHC * x."HC Grad."), x."Sw Hydrostatic")
forall x:"MASP Open Hole Key", x."TVDmud" = MAX(0, *(-(x."Open Hole Depth yielding highest MASP- RKB TVD (Ft)",
        x."Casing Section Key"."Well Data Key"."RKB-ML"), x."Mud Gradient Key"."Mud Fraction"))
forall x:"MASP Open Hole Key", x."TVDHC" = MAX(0, *(-(x."Open Hole Depth yielding highest MASP- RKB TVD (Ft)",
        x."Casing Section Key"."Well Data Key"."RKB-ML"), x."Mud Gradient Key"."Gas Fraction"))

forall x:"MASP Shoe Key", x."MASP Shoe" =
        -(-(-(x."Deepest Exposed Shoe Below This Shoe - RKB TVD (Ft)" * 
        x."Frac Gradient at Deepest Exposed Shoe (ppg)" * x.Constant),
        x.TVDmud * x.MW * x.Constant, x.TVDHC * x."HC Grad."), x."Sw Hydrostatic")
forall x:"MASP Shoe Key", x."MW" = x."OH Key"."Max Mud Weight at OH Depth (Downhole,ppg)"
forall x:"MASP Shoe Key", x."Frac Gradient at Deepest Exposed Shoe (ppg)" = 
        x."Exposed Shoe Key"."Frac Gradient at Deepest Exposed Shoe (ppg)"

forall x:"MASP Key", x.MASP = MIN(x."MASP Openhole", x."MASP Shoe")
forall x:"MASP Key", x."MASP Openhole" = x."MASP Open Hole Key"."MASP Open Hole" 
forall x:"MASP Key", x."MASP Shoe" = x."MASP Shoe Key"."MASP Shoe" 
		
forall x:"Exposed Shoe Key", x."Frac Gradient at Deepest Exposed Shoe (ppg)" = 
        x."PPFP Key"."Salt FG (Using OBG + 1000 psi (ppg)"

forall x:"Burst Calculation Key",   x.Constant=.052

    \end{verbatim}
     \end{footnotesize} \vspace*{-.3in} 
     
      \caption{Equations for Schema B}
    \label{fig:eqsB}
    \end{figure}
    
    \begin{figure}
    \begin{minipage}[t]{0.3\textwidth}
    \begin{center}
\includegraphics[width=1.8in]{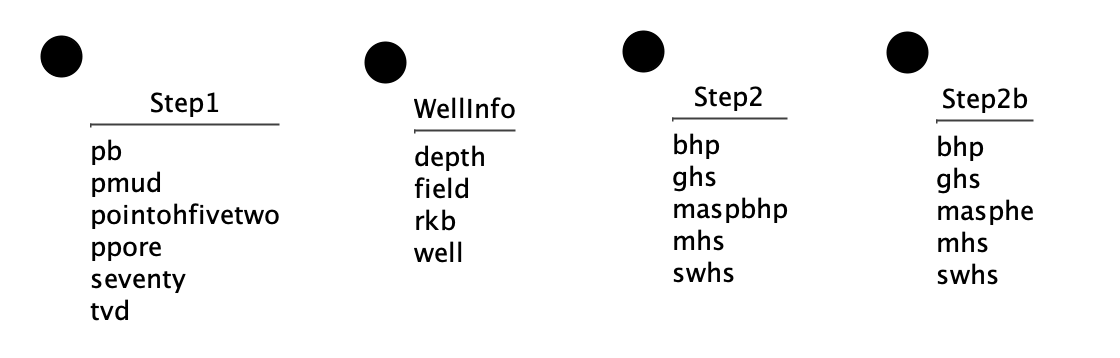}
    \end{center}
    \end{minipage}
    \begin{minipage}[t]{0.7\textwidth}
    \vspace{-.6in}
     \begin{footnotesize}
\begin{verbatim}
forall i:Step1, casingburst(i) = (i.seventy * i.pb) 
        - i.tvd * i.pointohfivetwo * (i.pmud - i.ppore)
forall i:Step2 , maspbhp (i) = i.bhp - (i.mhs + i.ghs + i.swhs)
forall i:Step2b, maspshoe(i) = i.bhp - (i.mhs + i.ghs + i.swhs)
    	\end{verbatim}
 	\end{footnotesize}
 	\end{minipage} \vspace*{-.2in}
\caption{Signature and Equations for Overlap Schema}
\label{fig:schemaO}
\end{figure}
\begin{figure}
\begin{footnotesize}
\begin{verbatim}
Step1 -> "MASP Calc. Step 1"                                pointohfivetwo -> lambda x, .052
pmud -> lambda x, x."Interval"."Downhole Mud Weight"        pb -> lambda x, x."Casing Section"."Burst Rating"
ppore -> lambda x, x."Zone Name"."Backup Pore Pressure"     tvd -> lambda x, x."RKB TVD"."RKB TVD"
casingburst -> lambda x, x."70% Burst (corrected)"          seventy -> lambda x, x."De-Rated Percent"

Step2 -> "MASP Calc. Step 2a"                   bhp -> lambda x, x."Pressure at Deepest OH Depth"
mhs -> lambda x, x."Mud Hydrostatic (BHP)"      ghs -> lambda x, x."Gas Hydrostatic (BHP)"
swhs -> lambda x, x."SW Hydrostatic"            maspbhp -> lambda x, x."MASP BHP"

Step2b -> "MASP Calc. Step 2a"                  bhp ->  lambda x, x."Frac Pressure at Deepest Shoe"
mhs ->  lambda x, x."Mud Hydrostatic (Shoe)"    ghs ->  lambda x, x."Gas Hydrostatic (Shoe)" 
swhs -> lambda x, x."SW Hydrostatic"            maspshoe ->  lambda x, x."MASP Shoe"

WellInfo -> "Header Info"       well -> lambda x, x."Well Name"
rkb -> lambda x, x."RKB-ML"     depth -> lambda x, x."WD"       field -> lambda x, x."Field"
\end{verbatim}
\end{footnotesize} \vspace*{-.2in}
\caption{Schema Mapping from Overlap to A}
\label{fig:mapA}
\end{figure}

\begin{figure}
\begin{footnotesize}
\begin{verbatim}
Step1 -> "Burst Calculation Key"        pointohfivetwo -> lambda x, .052        tvd -> lambda x, x.TVD
pb -> lambda x, x."Item of Interest"."Item of Interest - Burst Rating (psi)"
pmud -> lambda x, x.DHEMW               ppore -> lambda x, x."Backside EMW"     
casingburst -> lambda x, x."Burst Rating Corrected for MW & Backside"
seventy -> lambda x, x."Material Utilization Factor"

Step2 -> "MASP Key"     swhs -> lambda x, x."MASP Open Hole Key"."Sw Hydrostatic"
mhs -> lambda x, x."MASP Open Hole Key"."MW" * x."MASP Open Hole Key"."TVDmud" * x."MASP Open Hole Key"."Constant"
ghs -> lambda x, x."MASP Open Hole Key"."HC Grad." * x."MASP Open Hole Key"."TVDHC"
maspbhp -> lambda x, x."MASP Open Hole Key"."MASP Open Hole"
bhp -> lambda x, *(x."MASP Open Hole Key"."Pore Pressure at OH Depth (ppg)",
 		*(x."MASP Open Hole Key"."Constant", x."MASP Open Hole Key"."Open Hole Depth yielding highest MASP- RKB TVD (Ft)"))

Step2b -> "MASP Key"        swhs -> lambda x,  x."MASP Shoe Key"."Sw Hydrostatic"
mhs -> lambda x, x."MASP Shoe Key"."TVDmud" * x."MASP Shoe Key"."MW" * x."MASP Shoe Key"."Constant"
ghs -> lambda x, x."MASP Shoe Key"."HC Grad." * x."MASP Shoe Key"."TVDHC"
maspshoe -> lambda x,  x."MASP Shoe Key"."MASP Shoe"
bhp -> lambda x, *(*(x."MASP Shoe Key"."Deepest Exposed Shoe Below This Shoe - RKB TVD (Ft)", 
    x."MASP Shoe Key"."Frac Gradient at Deepest Exposed Shoe (ppg)"), x."MASP Shoe Key"."Constant")
    
WellInfo -> "Well Data Key"     well -> lambda x, x."Well"      rkb -> lambda x, x."RKB-ML"     
depth -> lambda x, x."Water Depth"      field -> lambda x, x."Field/Prospect"
\end{verbatim}
\end{footnotesize} \vspace*{-.2in}
\caption{Schema Mapping from Overlap to B}
\label{fig:mapB}
\end{figure}

    \begin{figure}
       \begin{footnotesize}
\begin{verbatim}
////////////////// source A /////////////////
forall bc:"MASP Calc. Step 1",  bc."Casing Section" =  bc."Zone Name"."Casing Section"
forall bc:"MASP Calc. Step 1",  bc."Casing Section".Interval = bc.Interval
forall bc:"MASP Calc. Step 1",	 bc.Interval.Well = bc."RKB TVD".Well
forall bc:"MASP Calc. Step 1",	 bc."Casing Section"."Total Vertical Depth".Well = bc."RKB TVD".Well
forall bc:"MASP Calc. Step 1", 	bc.Interval."Planned Section Total Depth".Well = bc."Zone Name"."RKB TVD".Well

forall x:"MASP Calc. Step 2a",	 x.Well = x."TVD Shoe".Well
forall x:"MASP Calc. Step 2a",	 x.Well = x."TVD Deepest OH".Well
forall x:"MASP Calc. Step 2a",	 x.Well = x."Interval"."Planned Section Total Depth".Well

////////////////// source B /////////////////
forall x:"Item of Inteerest Key", x."Liner Hangoff Key"."Top Casing Section / Nominal Casing - Size" = 
    x."Top Casing Section / Nominal Casing - Size"
 	
 forall x:"MASP Key", x."Casing Section Key" = x."MASP Open Hole Key"."Casing Section Key"
 forall x:"MASP Key", x."Casing Section Key" = x."MASP Open Hole Key"."Mud Gradient Key"."Casing Section Key"
 forall x:"MASP Key", x."Casing Section Key" = 
    x."MASP Open Hole Key"."OH Key"."Top Casing Section / Nominal Casing - Size"
 forall x:"MASP Key", x."MASP Open Hole Key"."Mud Gradient Key"."Casing Section Key" =
    x."MASP Shoe Key"."Casing Section Key"
 forall x:"MASP Key",	x."MASP Shoe Key"."Exposed Shoe Key"."Top Casing Section / Nominal Casing - Size" = 
    x."MASP Shoe Key"."Casing Section Key"
 forall x:"MASP Key",	x."MASP Open Hole Key"."OH Key"."Top Casing Section / Nominal Casing - Size" =
     x."MASP Shoe Key"."Casing Section Key"
     
////////////////// both sources /////////////////
forall x:"Interval Info", x.Interval.Well."Well Name" = 
    x."Item of Interest"."Top Casing Section / Nominal Casing - Size"."Well Data Key"."Well"

forall x:"MASP Calc. Step 2a", x."MASP Open Hole Key"."Mud Gradient Key"."Casing Section Key"."Well Data Key" = 
    x."MASP Shoe Key"."OH Key"."Top Casing Section / Nominal Casing - Size"."Well Data Key"

forall x:"MASP Calc. Step 1", x."Item of Interest"."Liner Hangoff Key"."Top Casing Section / Nominal Casing - Size"
    ."Well Data Key" = x."Casing Section"."Interval".Well

forall x:"MASP Calc. Step 1", x."Item of Interest"."Liner Hangoff Key"."Top Casing Section / Nominal Casing - Size"
    ."Well Data Key" = x."Zone Name"."Casing Section"."Interval".Well

forall x:"MASP Calc. Step 1", x."Item of Interest"."Liner Hangoff Key"."Top Casing Section / Nominal Casing - Size"
    ."Well Data Key" = x."Interval".Well

forall x:"MASP Calc. Step 1", x."Item of Interest"."Liner Hangoff Key"."Top Casing Section / Nominal Casing - Size"
    ."Well Data Key" = x."Zone Name"."Casing Section"."Interval".Well

forall x:"MASP Calc. Step 2a", x.Interval.Well = 
    x."MASP Open Hole Key"."Mud Gradient Key"."Casing Section Key"."Well Data Key"

forall x:"MASP Calc. Step 2a", x.Well = 
    x."MASP Open Hole Key"."Mud Gradient Key"."Casing Section Key"."Well Data Key"

forall x:"Item of Inteerest Key", x."Top Casing Section / Nominal Casing - Size"."Well Data Key" = 
    x."Liner Hangoff Key"."Top Casing Section / Nominal Casing - Size"."Well Data Key"
 	\end{verbatim}
 	\end{footnotesize}
 	\caption{Additional Equations for Better Data Integration}
 	\label{fig:eqsMore}
 	\end{figure}
 	
    \subsubsection{Overlap Schema and Mappings}
    
\begin{figure}
 	    \centering
 	    \includegraphics[width=3in]{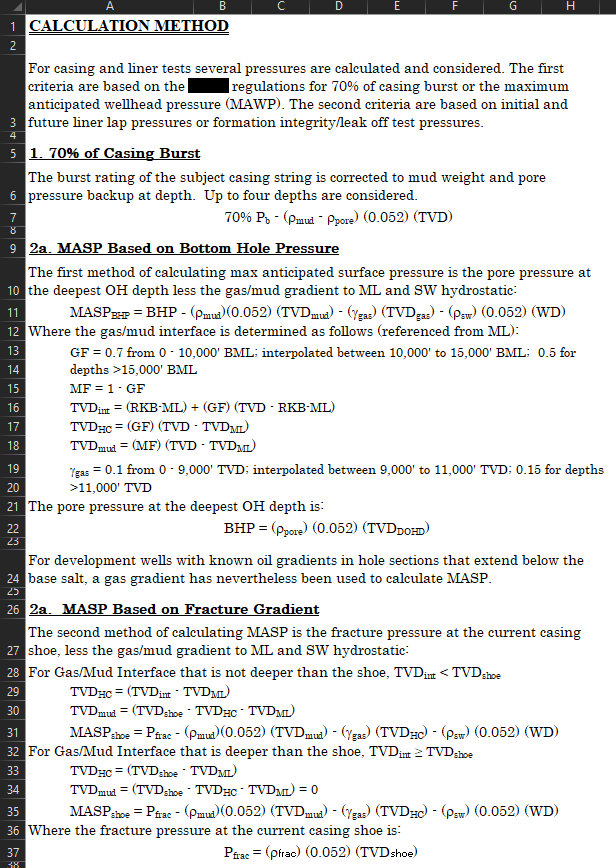}
 	    \caption{MASP Definition as Overlap}
 	    \label{fig:maspreg}
 	\end{figure}

    The signature (sorts and symbols) of schema $O$, the overlap schema, are defined in Figure~\ref{fig:schemaO}, along with its equations. The overlap schema is inspired by the English definition of the MASP calculation included with the original Excel sheet, shown in Figure~\ref{fig:maspreg}.
    
    The three equations in the overlap schema generate the verification conditions described in section~\ref{sec:vc}; our methodology checks that the equations in the overlap schema are entailed by the equations of each source schema.  In this way, we need not appeal to the authors of the source schemas; we can check if the equations (semantics) we desire (in the overlap) are already present in the source schemas.  It is for this reason we describe our methodology as ``consensus-free''.
    
    The schema mappings from the overlap schema to source A's schema and source B's schema are shown in Figures~\ref{fig:mapA} and \ref{fig:mapB} respectively.  As required by algebraic data integration, for each overlap table, we give a table in source A (or B, respectively), and for each overlap column, we give an expression in source A (or B, respectively).    
    
    The four tables that make up the overlap schema are defined as follows. The ``WellInfo'' table captures general information about the subject well to allow the input models to integrate data for calculations being performed on the same well. The first step of the MASP calculation, defined in the ``Step1'' table, represents the de-rated burst pressure rating of the subject casing strings after being corrected for mud weight and pore pressure backup values at the depth of interest. This is a requirement based on government regulations, which stipulate that the subject casing string should be de-rated to 70 percent of the casing burst pressure rating. The next requirement in the MASP calculation determines the maximum anticipated surface pressure. This is performed for both bottomhole pressure and fracture gradient scenarios and is defined by the ``Step2'' and ``Step2b'' tables, respectively. The mappings from this overlap model to schema A and schema B define columns where each component of the overlap computation is contained within the input models.  
    
    These tables were chosen as overlap, and the schema mappings designed as they are, because the overlapping tables represent engineering requirements shared by many business units and engineering teams around the world. Mathematically verifying that an expression of these requirements is entailed by the equations in each source schema opens a door to new forms of well design assurance and improving the quality and cycle time of well engineering activities, as discussed in the conclusion.

\subsection{Suggesting rules}

For large spreadsheets, it is impractical for users to start from a blank screen and simply type hundreds or thousands of formulae to relate them.  Therefore data integration tools, including ours, provide heuristic suggestions as a starting point.  Rules that well names are preserved as foreign keys are traversed are good examples of rules that are easy to guess based on column names and on data.  The paper~\cite{schultz:hal-01767471} describes some heuristic rule suggestion algorithms for our particular algebraic data model, and many traditional heuristic algorithms~\cite{doan2012principles} can be re-used off-the-shelf in our system.

    \subsubsection{Result}
    
The result of schema integration can be easily described: we start by taking the disjoint union $S_A+S_B$ of the source schemas; then, for every table $o$ in the overlap schema $O$, we merge tables $M_A(o)$ and $M_B(o)$ and add equations $M_A(f)$ = $M_B(f)$ for every $f$ with domain $o$, ultimately resulting in a ``colimit''~\cite{schultz_wisnesky_2017} schema (theory) denoted $S_A +_{O,M_A,M_B} S_B$.  This schema is large and somewhat redundant in our case study because we are only merging three tables (``step 1'', ``step 2a'', and ``header/well info''), so we do not display it here.  However, in Figure~\ref{fig:eqsMore} we display additional equations that we added to the colimit schema to aid data quality during data integration-- arguably, some of these equations were forgotten in the original schemas.  Additionally, the integrated schema can be read off the integrated data from the full result in the appendix.

{\bf Remark.} The result of schema integration is only defined up to unique isomorphism, and so for example the names of tables in the result, and even the number of columns and equations in it, is not determined uniquely-- the CQL software has many options to allow users to select which resulting schema (theory) they want from among many choices.  In fact, under mild assumptions it is possible to prove that there is no ``best'' table and column naming scheme.  In practice, the integrated schema is often customized with user-specific naming of tables and columns.  

\subsection{Data Integration and Exchange}

We continue to follow the algebraic data integration pattern from Figure~\ref{fig:algint} and in this section define the data sources and their overlap that complete the pattern started in the previous section.   We now walk through the data contained in each table below.

    \subsubsection{Data A}
    
\begin{figure}
        \centering
        \vspace*{-.59in}
     \includegraphics[width=5.5in]{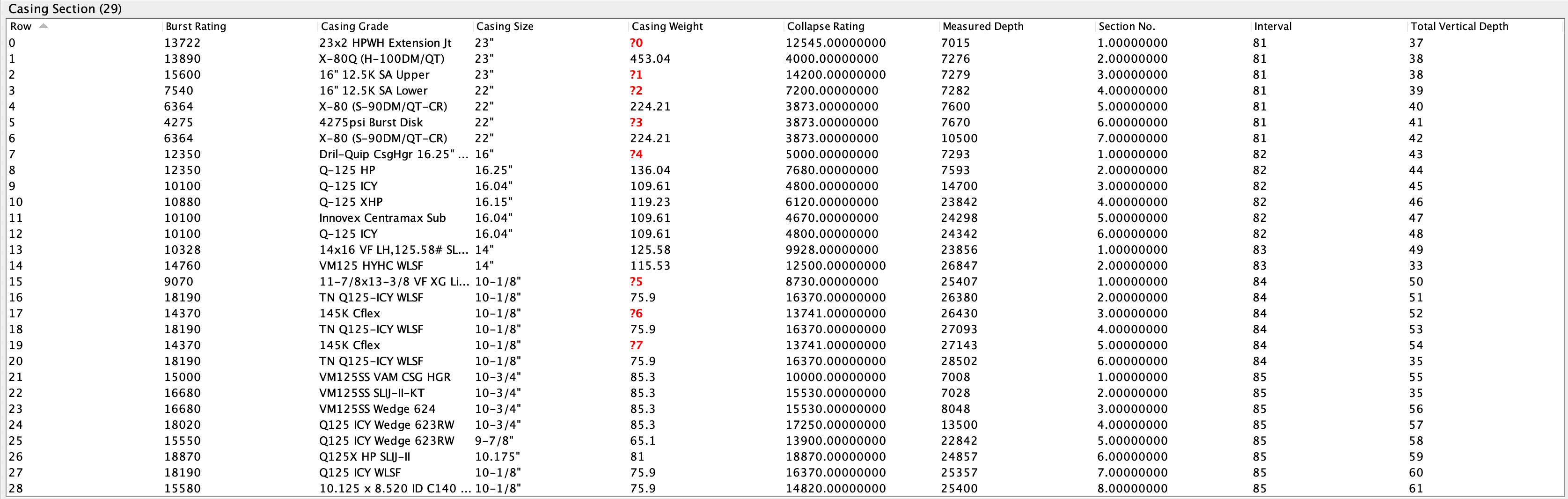}
    \includegraphics[width=5.5in]{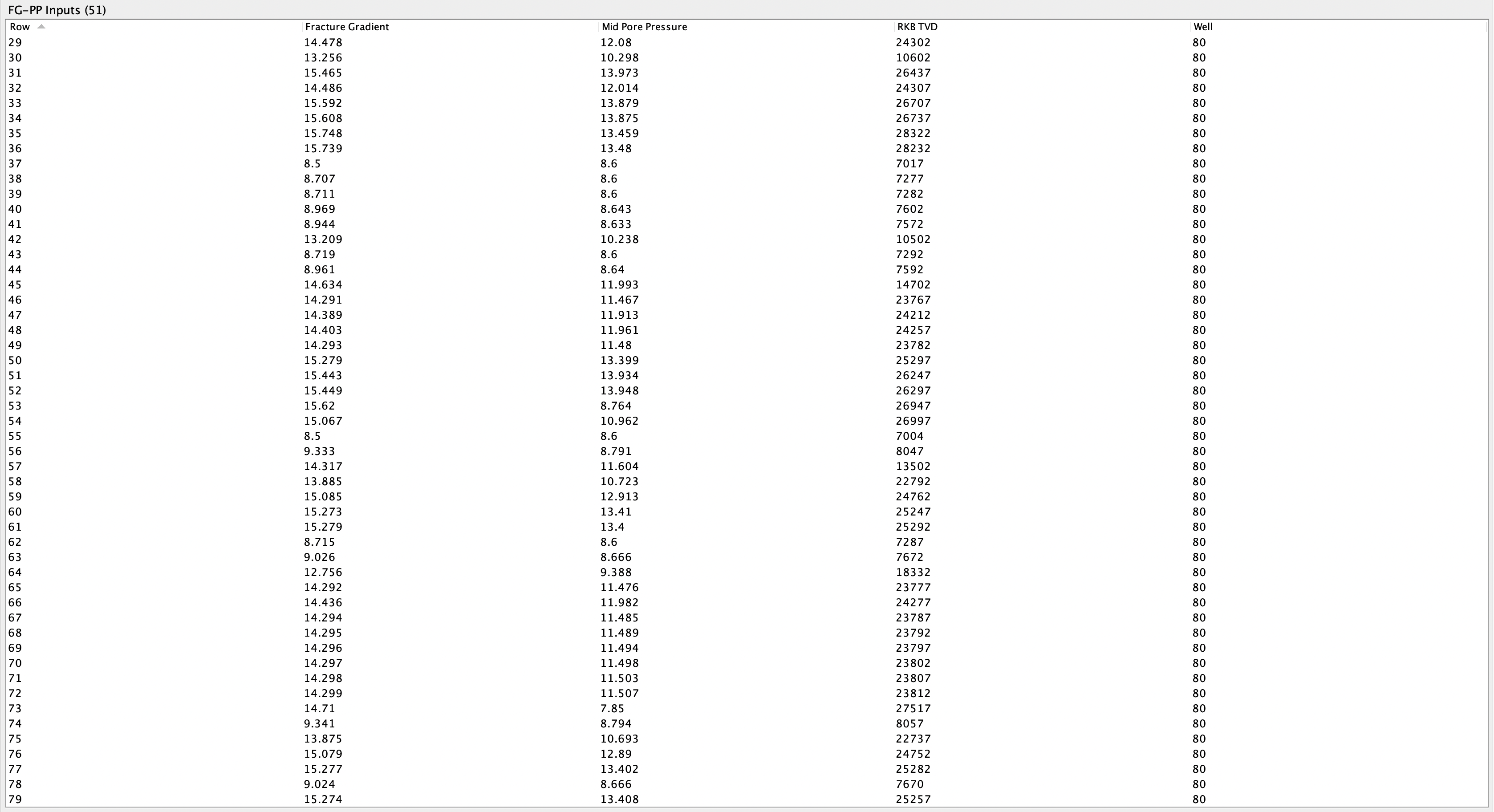}
        \includegraphics[width=5.5in]{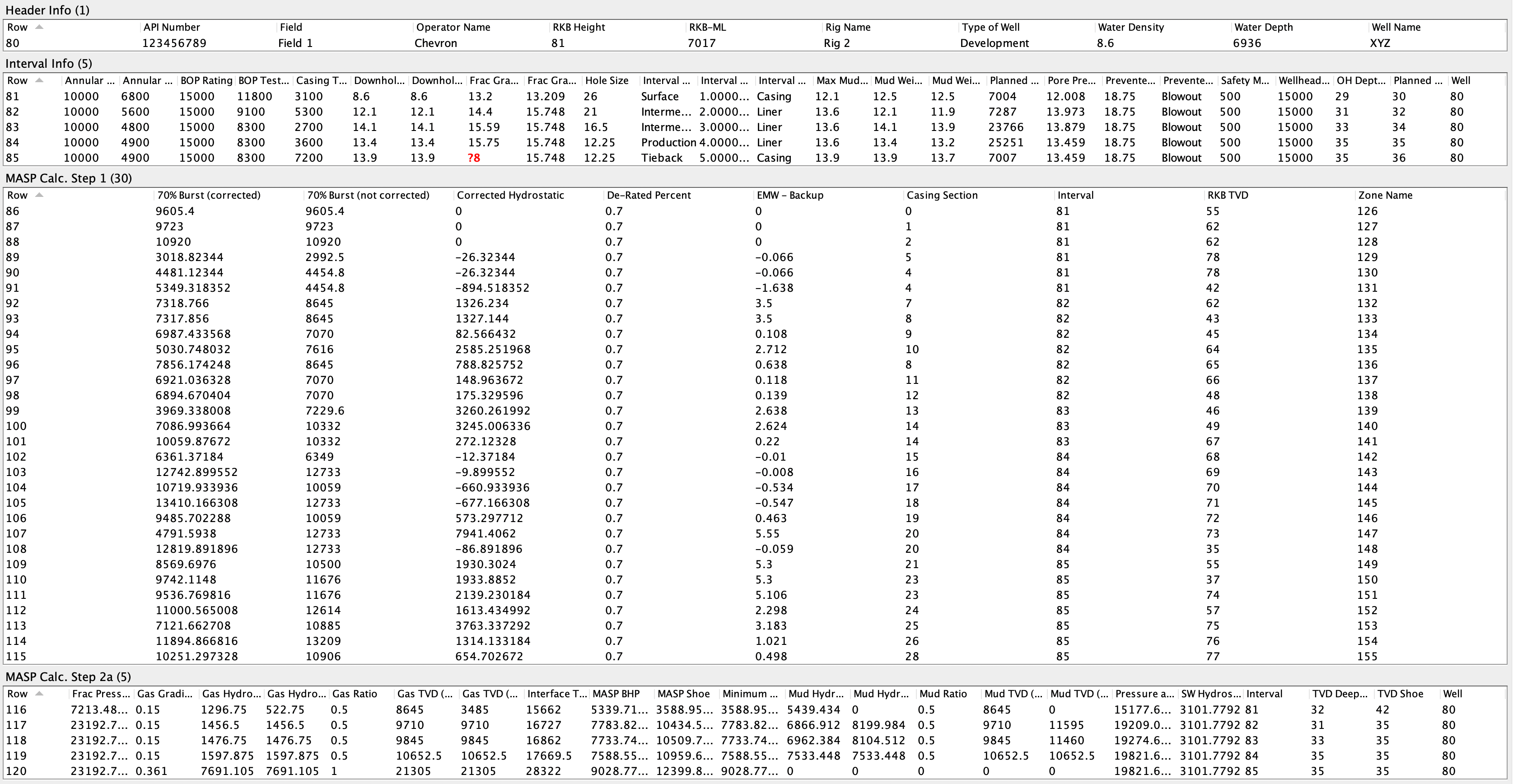}
    \includegraphics[width=5.5in]{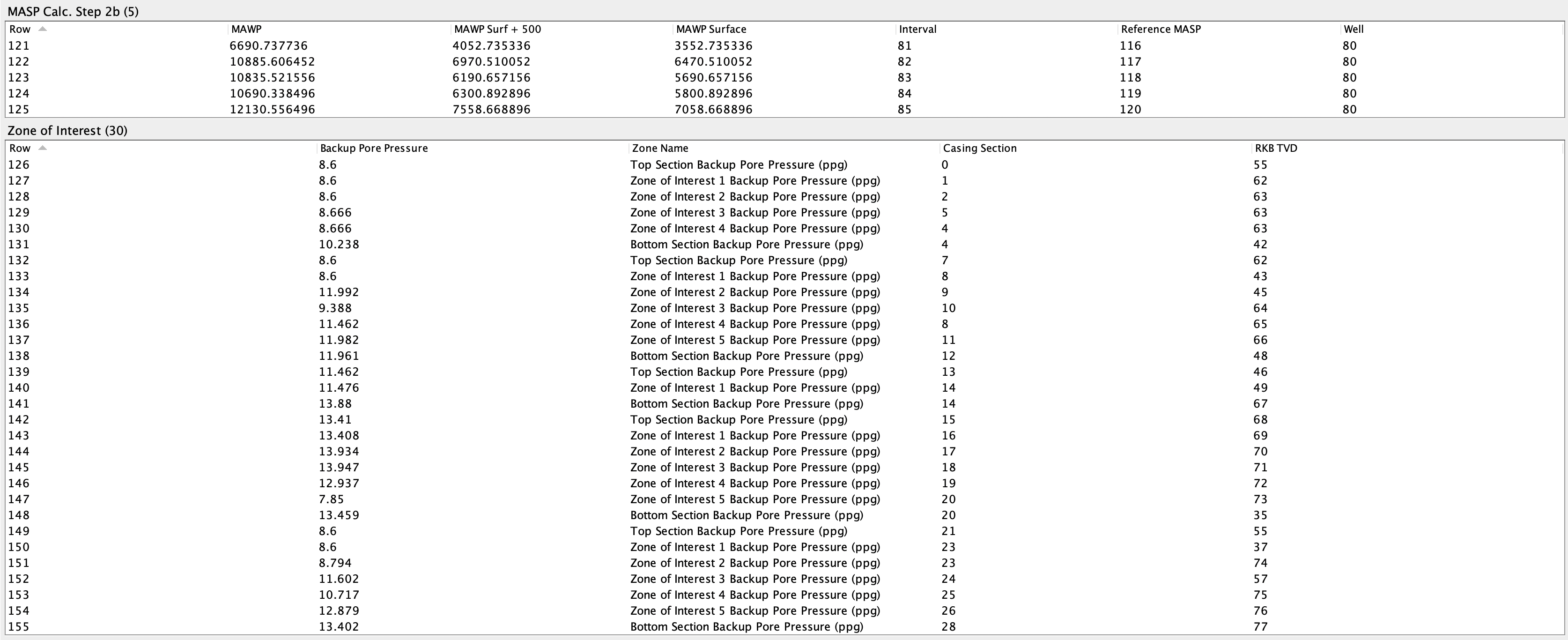}

        \caption{Data for Schema A}
        \label{fig:dataA}
    \end{figure}

      Input data for source A is shown in Figure~\ref{fig:dataA}. 
      
      
\begin{itemize}
    
\item    Casing Section Table
    
    \begin{itemize}
    \item Multiple casing intervals are included (e.g. 22", 16", etc).
    \item Question marks in the Casing Weight column reflect missing values in the original Excel MASP spreadsheet.
\end{itemize}

\item     FG-PP Inputs Table
    
    \begin{itemize}
    \item Over 4,000 rows were originally included in this table's sheet, but not all are imported into CQL for expediency. 
    \item This table represents geologic inputs from cross-domain stakeholders who govern the reservoir characteristics and Earth model.
\end{itemize}

\item    Header Info Table
    
    \begin{itemize}
    \item Only one well was modeled, so there is only one row.
\end{itemize}

\item     Interval Info Table
    
    \begin{itemize}
    \item Five rows reflect the casing/hole sections that are included.
    \item The question mark in the table reflects a missing fracture gradient value in the original Excel MASP spreadsheet.
\end{itemize}

 \item    MASP Calc. Step 1 Table
    
    \begin{itemize}
    \item The first criteria for calculating MASP is based on government regulations and includes calculating 70 percent of the burst pressure rating for each casing string. 
    \item This step also considers the burst pressure rating of the subject casing string when corrected to mud weight and pore pressure backup at various depths or zones of interest.
\end{itemize}

\item     MASP Calc. Step 2a Table
    
    \begin{itemize}
    \item Five rows reflect the MASP result (for each casing section) based on bottom hole pressure, which is calculated using the deepest open hole depth.
\end{itemize}

\item     MASP Calc. Step 2b Table
    
    \begin{itemize}
    \item Five rows reflect the MASP result (for each casing section) based on the fracture gradient, which is calculated using the depth of the deepest exposed casing string (known as the casing shoe).
\end{itemize}
  
\item    Zone of Interest Table
    
    \begin{itemize}
    \item This table reflects key attributes associated with each casing string, the components that make up that casing string, and the associated geological characteristics at a particular depth.
\end{itemize}

\end{itemize}

    \subsubsection{Data B}
    
    \begin{figure}[t]
        \centering
        \includegraphics[width=6.5in]{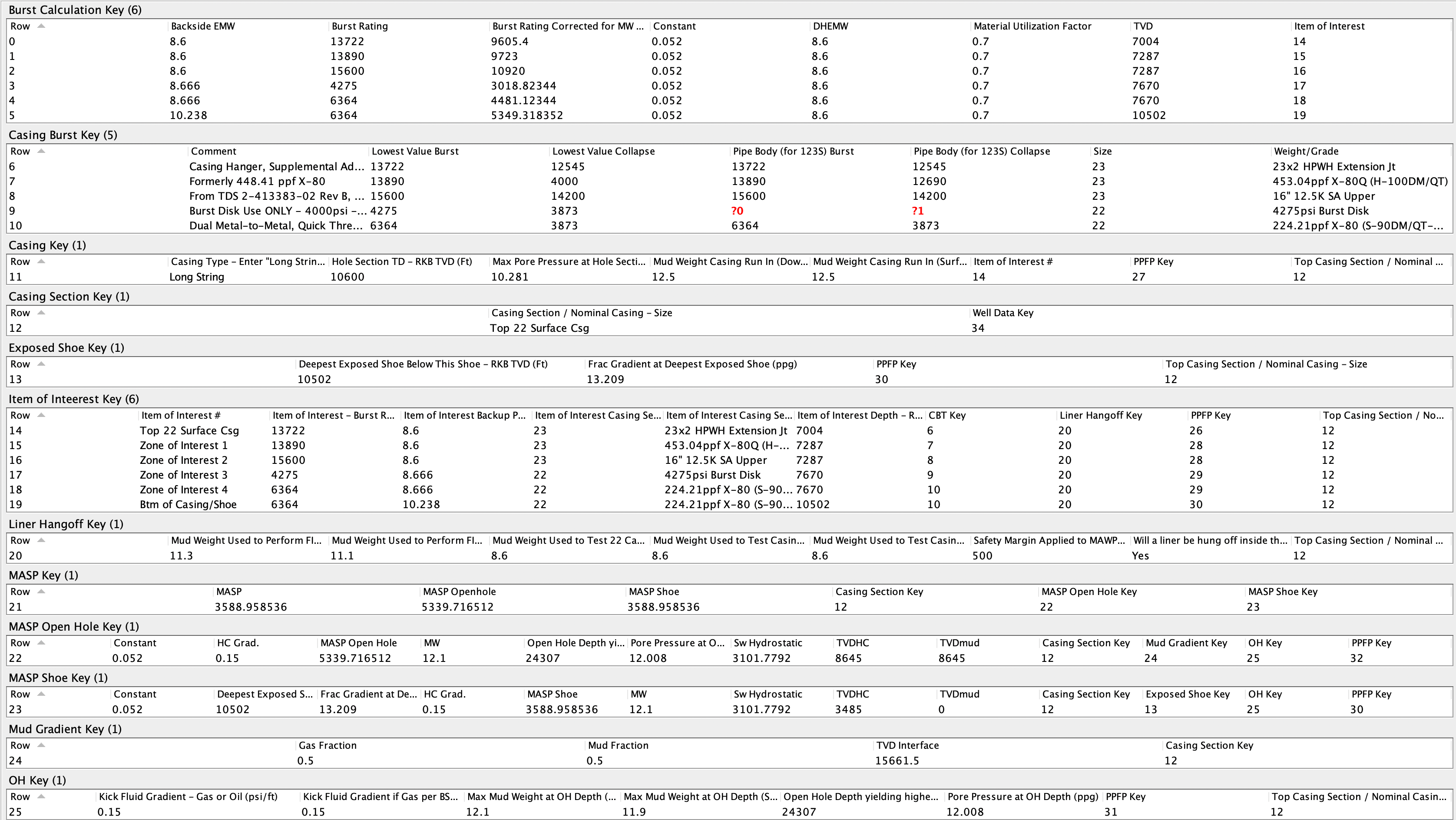}
        \includegraphics[width=6.5in]{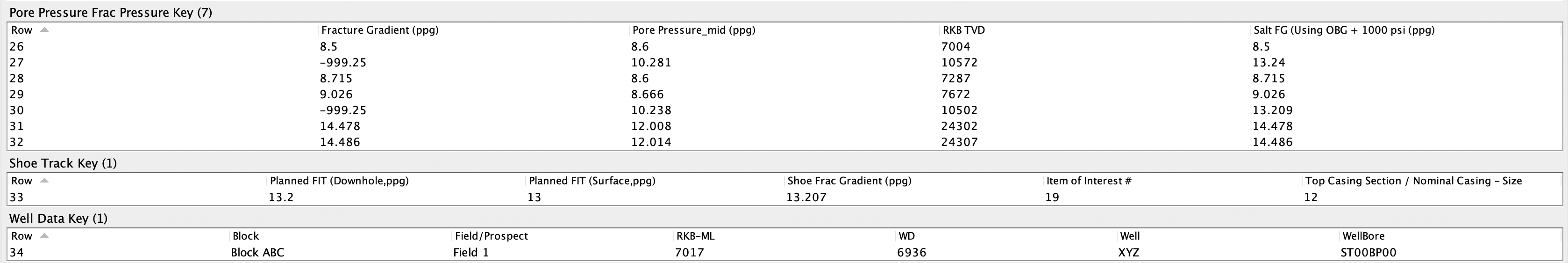}
        \caption{Data for Schema B}
        \label{fig:dataB}
    \end{figure}

The input data for source B is shown in Figure~\ref{fig:dataB}.  
B's data includes the same source data as A, except the MASP calculation is only performed on a single casing section (as opposed to five). As we will see later, this is the primary reason why B's data will be extended significantly in the case study's resulting data exchange, but A's data will be unchanged. Further, both A and B include ancillary columns from the source sheet that were not used in the MASP calculation, resulting in additional differences at the data level.  We now walk through the data contained in each table below.


\begin{itemize}
    
\item    Burst Calculation Key Table
    
    \begin{itemize}
    \item This table reflects the depths of interest considered for the MASP calculation of the 22" casing section.
    \item Key variables for step 1 of the MASP calculation (i.e., 70 percent burst pressure rating of the casing, de-rated and corrected for backup) are also included.
\end{itemize}

\item     Casing Burst Key Table
    
    \begin{itemize}
    \item This table contains casing specifications for the 22" hole section.
    \item Source B captured this data from a separate tab in the original Excel MASP sheet, whereas Source A did not use this table at all for casing data.
    \item Question marks represent missing values in the original Excel MASP sheet.
\end{itemize}

\item    Casing Key Table
    
    \begin{itemize}
    \item Only one well is modeled, hence there is only one row.
\end{itemize}

\item     Casing Section Key Table
    
    \begin{itemize}
    \item This table contains a basic description of the 22" surface casing.
\end{itemize}

 \item    Exposed Shoe Key Table
    
    \begin{itemize}
    \item This table informs the fracture gradient at the deepest exposed shoe, which is a key input for the MASP shoe calculation.
\end{itemize}

\item     Item of Intereest [sic] Table
    
    \begin{itemize}
    \item Similar to the Burst Calculation Key table, this table reflects the depths of interest where the MASP calculations were performed.
\end{itemize}

\item    Liner Hangoff Key Table
    
    \begin{itemize}
    \item Data for this table is included in anticipation of MASP calculation steps beyond the scope of the case study. So while the data are not used for step 1 or step 2 for MASP as defined in Figure~\ref{fig:maspreg}, they are typically used in subsequent steps of the casing design analysis.
\end{itemize}
  
\item    MASP Key Table
    
    \begin{itemize}
    \item This table contains a summary of results from the MASP open hole and MASP shoe calculations.
\end{itemize}

\item    MASP Open Hole Key Table
    
    \begin{itemize}
    \item This table contains the key inputs and calculations for the MASP open hole step of the analysis.
\end{itemize}

\item    MASP Shoe Key Table
    
    \begin{itemize}
    \item This table contains the key inputs and calculations for the MASP shoe step of the analysis.
\end{itemize}

\item    Mud Gradient Key Table
    
    \begin{itemize}
    \item This table informs the fluid column scenario for the MASP calculation (i.e., the ratio of gas-to-mud in the wellbore).
\end{itemize}

\item    OH Key Table
    
    \begin{itemize}
    \item This table contains key information for the MASP open hole scenario, such as pore pressures and mud weights.
\end{itemize}

\item   Pore Pressure Frac Pressure Key Table
    
    \begin{itemize}
    \item The original source table contains over 4,000 rows of data, but not all are imported into CQL for expediency.
    \item This table represents geologic inputs from cross-domain stakeholders who govern the reservoir characteristics and earth model.
\end{itemize}

\item    Shoe Track Key Table
    
    \begin{itemize}
    \item This table contains information for MASP shoe scenario, specifically the fracture gradient at the shoe depth. 
\end{itemize}

\item    Well Data Key Table
    
    \begin{itemize}
    \item Only one well is modeled, so there is a single row of data reflecting the general well header information.
\end{itemize}

\end{itemize}

    \subsubsection{Overlap Data and Mappings}

Although the algebraic data integration design pattern requires  an overlap instance $I_O$ on the overlap schema $S_O$, it can be unfeasible to generate such instances by hand, because they are often large and difficult to describe as models.  So, instead we describe rules--Horn clauses; implications of equations--that implicitly specify the overlap, including the required data mappings out of the overlap data.  Figure~\ref{fig:merge} states that step 1 rows, now coming from both source A and source B, should be merged when they agree on two particular columns, and similarly for step 1a.  If such a (recursive) merge is contradictory, CQL reports this, as described in Section~\ref{sec:vc}.  

{\bf Remark.}  Not all sets of Horn clauses (implications of equations) induce an overlap instance and two data mappings as required for algebraic integration.  If we were to follow the algebraic integration pattern precisely~\cite{schultz_wisnesky_2017},  we would instead give a ``query'' of a specific form that materializes the matching pairs of source A rows and source B rows as an overlap instance.  Such a query is a re-statement of our Horn clauses.    

{\bf Remark.}  As an alternative to thinking of row merge rules as specifying an overlap instance for which a colimit is taken, we can also think of them as specifying a ``lifting problem'' that we are solving.  If we do this, as occurs often in practice, then we can generalize our rule language to include existential quantifiers on the right-hand side of implications~\cite{schultz_wisnesky_2017}, dramatically increasing their expressive power.

\begin{figure}
\begin{footnotesize}
\begin{verbatim}
forall x y : "MASP Calc. Step 1", x."70% Burst (corrected)" = y."Burst Rating Corrected for MW & Backside" -> x = y
forall x y : "MASP Calc. Step 2a", x."MASP BHP" = y."MASP Openhole" -> x = y    
forall x y : "Header Info", x."Well Name" = y."Well Name" -> x = y
forall x y : "Header Info", x."Well" = y."Well" -> x = y
\end{verbatim}
\end{footnotesize}

        \caption{Row merge rules}
        \label{fig:merge}
    \end{figure}
    
    \begin{figure}
\includegraphics[width=6.5in]{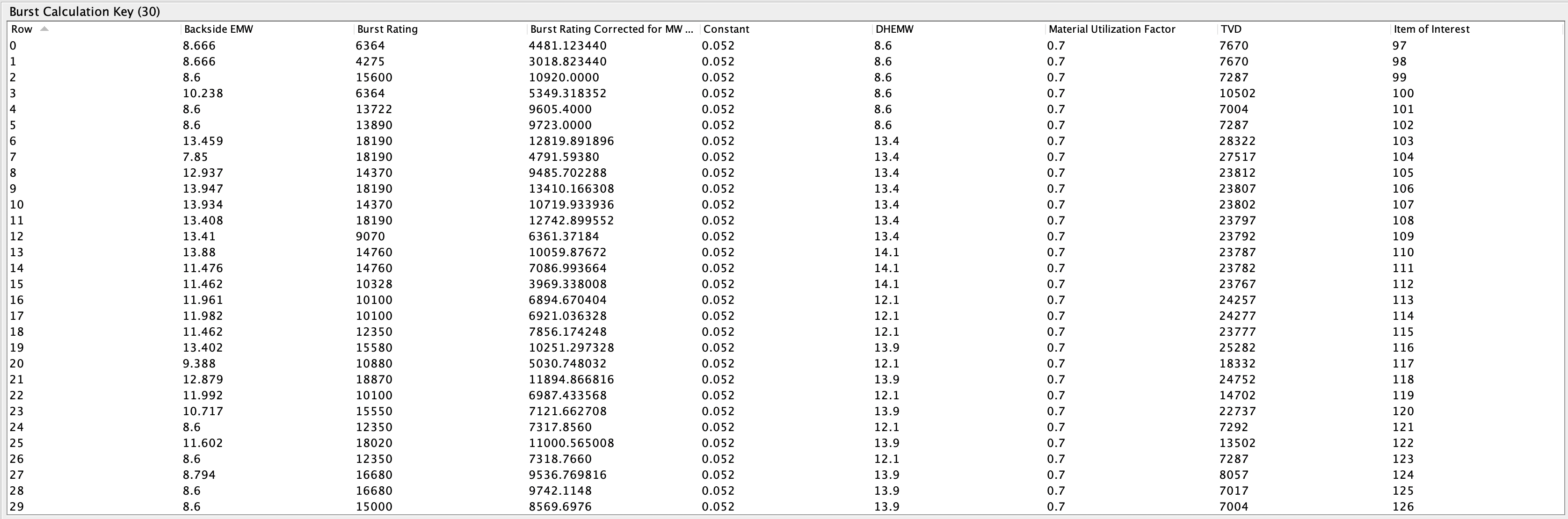}
\includegraphics[width=6.5in]{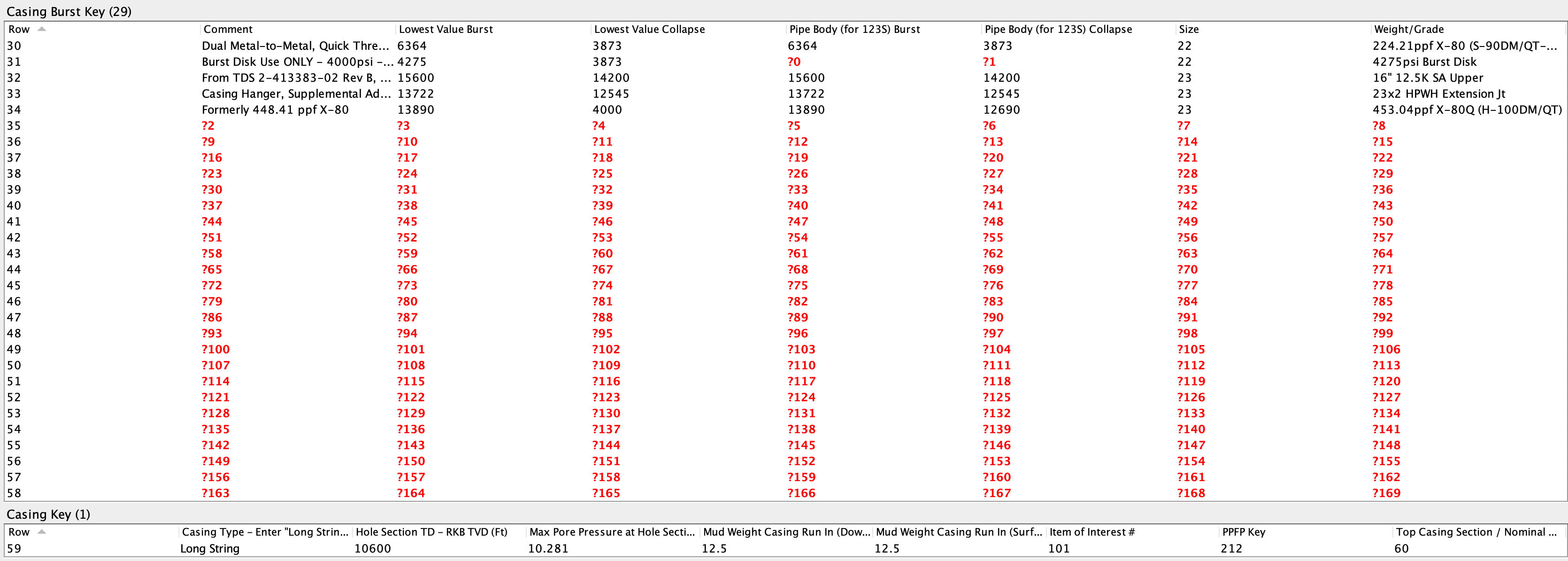}
\includegraphics[width=6.5in]{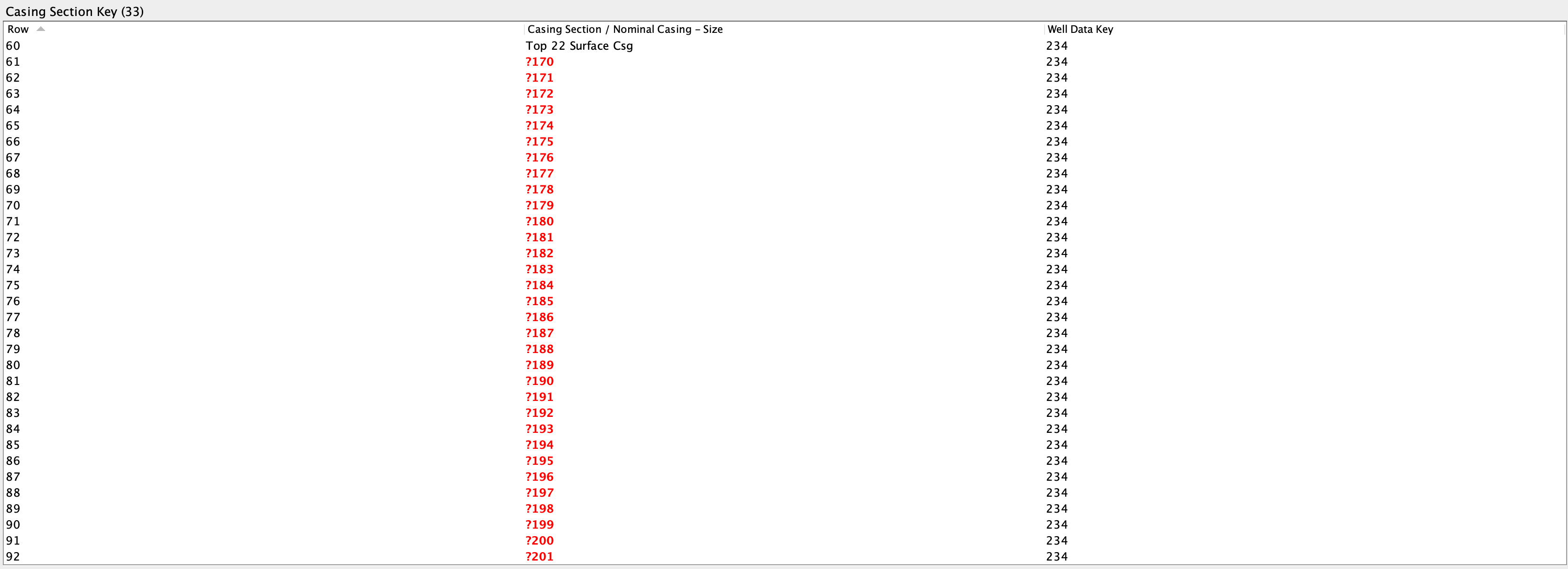}

        \caption{Data Exchange of Source A to Source B, 1 of 3}
        \label{fig:outA}
    \end{figure}
    
     \begin{figure}
\includegraphics[width=6.5in]{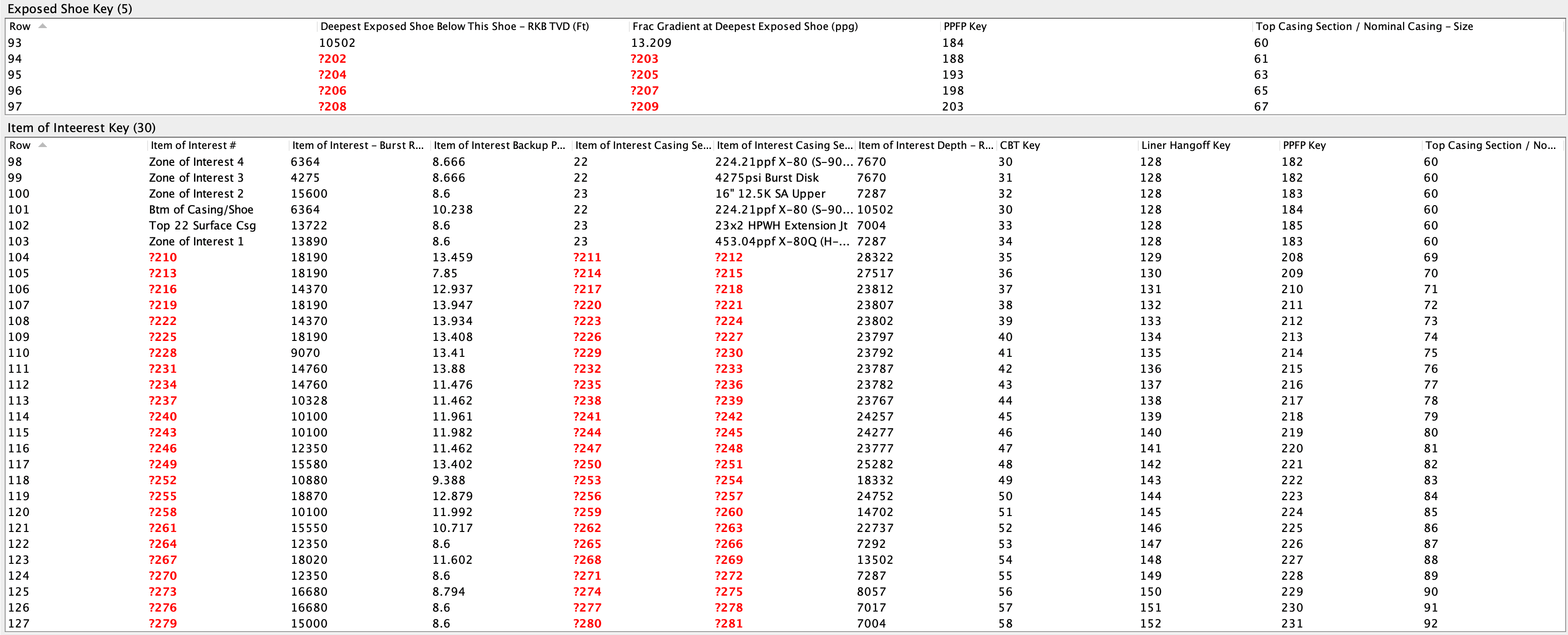}
\includegraphics[width=6.5in]{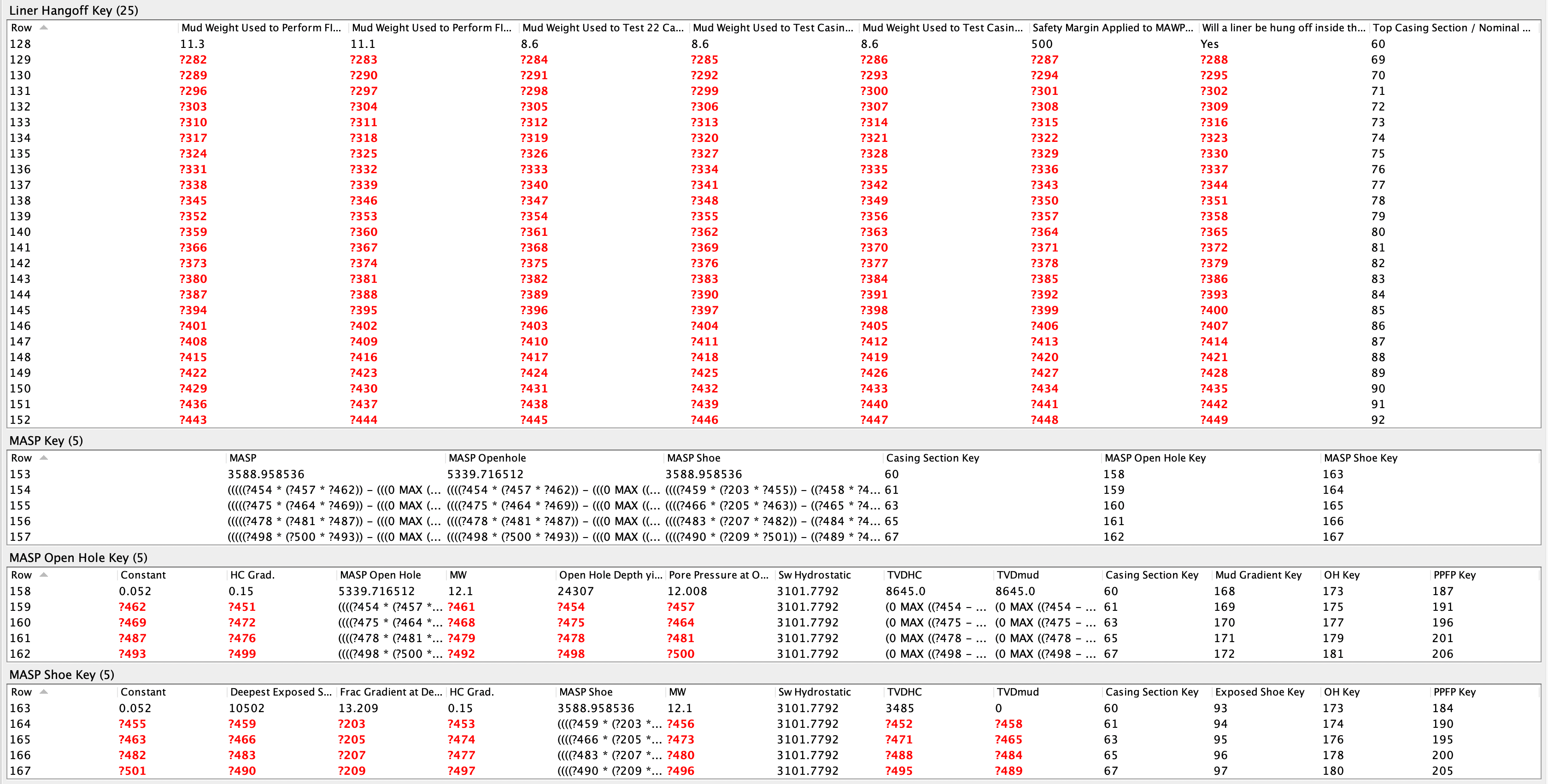}
\includegraphics[width=6.5in]{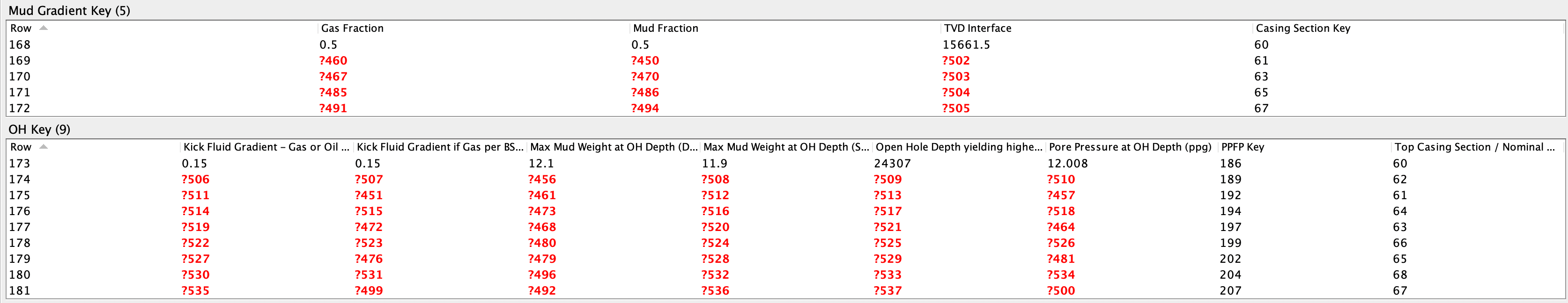}

        \caption{Data Exchange of Source A to Source B, 2 of 3}
        \label{fig:outB}
    \end{figure}
    
     \begin{figure}
\includegraphics[width=6.5in]{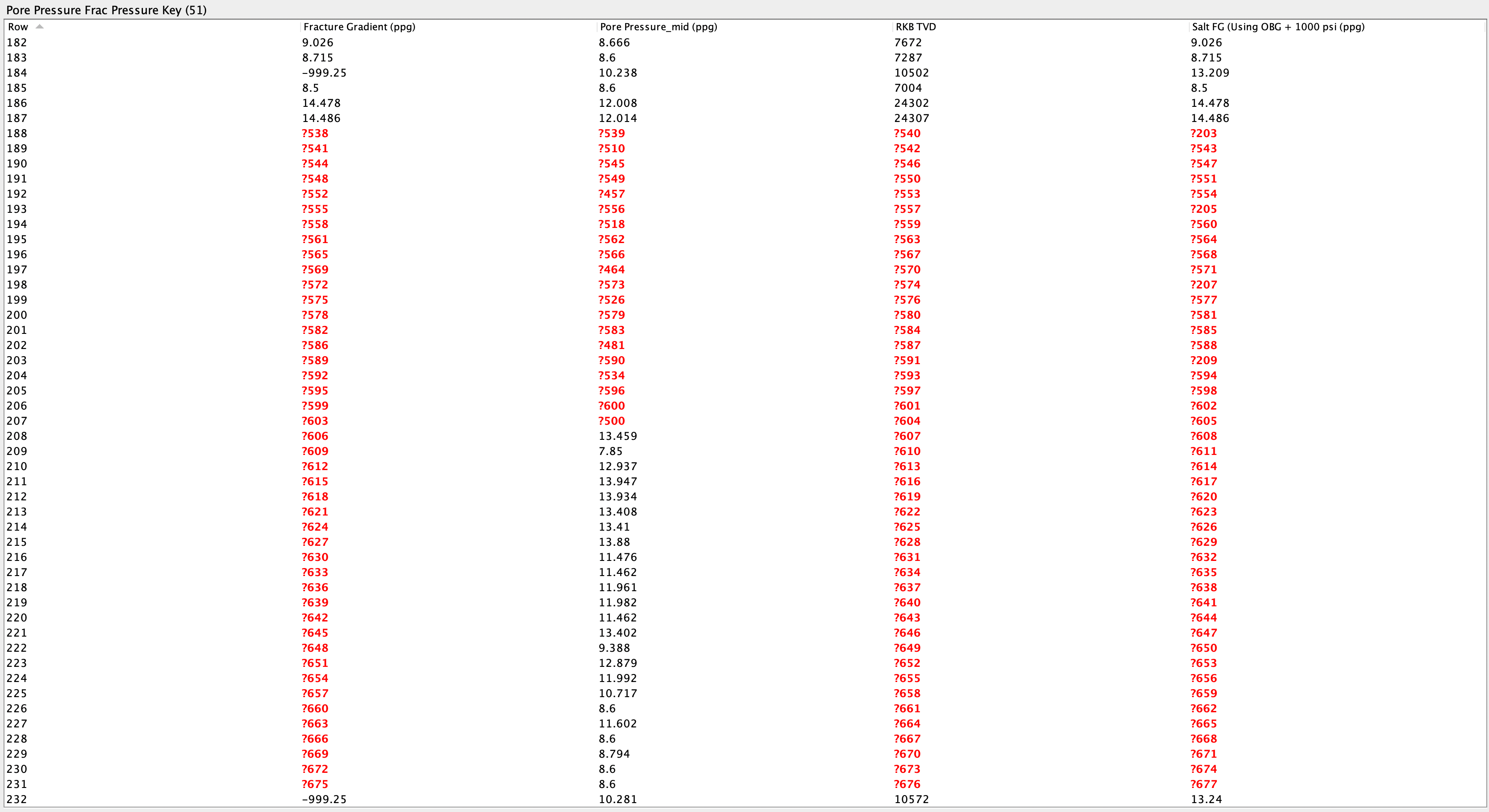}
\includegraphics[width=6.5in]{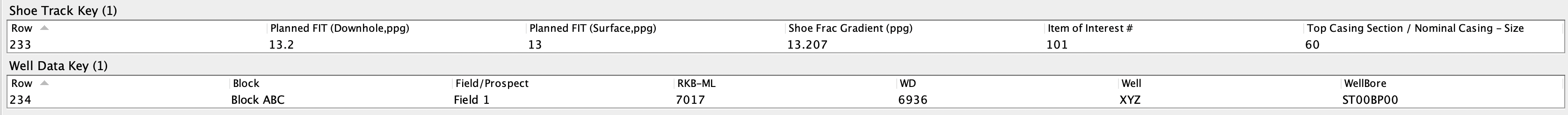}

        \caption{Data Exchange of Source A to Source B, 3 of 3}
        \label{fig:outC}
    \end{figure}

    \subsubsection{Result}
    
    The the integrated result in algebraic data integration is given by taking a colimit of the input databases, each suitably pushed forward onto the colimit schema (plus additional user-defined equations, shown in Figure~\ref{fig:eqsMore}) via ``left Kan extension''~\cite{schultz_wisnesky_2017}.  However, because this integrated result subsumes both spreadsheets, it is too large to conveniently display here (we do so in the appendix).  So instead, we will focus on data exchange between the source sheets.  That is, we project the integrated result back on to each source and analyze the difference with the original source.  When we do this on this example, we find that source A is unchanged, but source B is extended significantly.  Sometimes, the extensions to source B are with concrete values, such as in the ``Burst Calculation Key'' table, but other times the extensions are with blank values (labeled nulls), indicating that the source B is incomplete with respect to source A as defined by the overlap and mappings.  Intuitively, this phenomenon happens because source B is in some sense contained source A, which has more rows to begin with. The ``round-tripped'' B data is displayed in Figures~\ref{fig:outA} and \ref{fig:outB}.  The source owners and subject matter experts made the following observations about each table after the data exchange:
    
\begin{itemize}  

\item Burst Calculation Key Table
    
    \begin{itemize}
    \item Whereas source B started with six rows in this table, the data exchange with source A through the integrated schema results in 30 rows of data. This is due to source B sharing six overlapping rows with source A (these were merged together) while receiving 24 rows as new information gain.
    \item The specific rows being merged are based on the 70 percent burst rating (corrected) values for each source.
    \item Filled-in values can be traced primarily to the MASP ``Calc. Step 1'' table from source A, but includes other  information implied by myriad equations and mappings.
\end{itemize}

\item Casing Burst Key Table
    
    \begin{itemize}
    \item This table is primarily associated with the ``Casing Section'' table from source A. This is observed by the 29 rows in both tables.
    \item The question marks are due to the fact that source A does not include any information about these columns (such as ``Lowest Value Collapse''). However, it was observed while creating rules that the source B Size column did not find a match in source A because of a difference in data type. This casing size data is represented as a string in source A and as a float in source B, where in particular, where source B  writes 22 where source A writes 22''.  To merge along such attributes, we can add float to string and string concatenation functions to our ambient ``theory of excel'' and then write an equation using these new primitives (e.g. 22''=concat(floatToString(22),'').
    \item From source A's perspective, this table's columns are not necessary in the MASP calculation, so they are largely excluded from data exchange.  Instead, source B includes these values because source B's data was pulled from a separate dataset containing casing specification information. 
\end{itemize}

\item  Casing Key Table
  
    \begin{itemize}
    \item Most of this data is not used for the MASP calculation and is in source B and not source A, explaining why there is only one row for source B.
\end{itemize}

\item  Casing Section Key Table
    
    \begin{itemize}
    \item Source B only models one string of casing (the 22" casing section), which is reflected in the fully populated row. The remaining rows are from source A and reflect the additional casing sections/components that are transferred from the original MASP sheet.
\end{itemize}

\item  Exposed Shoe Key Table
    
    \begin{itemize}
    \item Source B gains four rows of data because the MASP calculation is performed for five hole sections for source A. 
\end{itemize}

\item Item of Intereest Key Table [sic]
    
    \begin{itemize}
    \item Like the ``Burst Calculation Key'' table, the ``Item of Intereest Key'' table begina with six rows of data and, through the data exchange with aource A, resulta in an additional 24 rows of new data. However, in this case, there are additional rules in place to infer data values for two additional columns (``Item of Interest - Burst Rating (psi)'' and ``Item of Interest Depth - RKB TVD (Ft)''.
\end{itemize}

\item  Liner Hangoff Key Table
    
    \begin{itemize}
    \item This table is not used for the MASP calculation, and the only row of data comes from Source B. 
\end{itemize}

\item  MASP Key, MASP Open Hole Key, MASP Shoe Key, and Mud Gradient Key Tables
    
    \begin{itemize}
    \item For these tables, source B gains four rows of data each because the MASP calculation is performed for five hole sections for source A. 
\end{itemize}

\item  OH Key Table
    
    \begin{itemize}
    \item Most of the data from this table is not used for the MASP calculation, so there are minimal opportunities to create merge rules.  
\end{itemize}

\item  Pore Pressure Frac Pressure Key Table
    
    \begin{itemize}
    \item This table matches with the sources on six rows because six pore pressure / fracture gradient inputs were used in the 22" casing section MASP calculated by source B. Because this data is shared by the subsurface discipline, and source A, and source B, the entire table could, in principle, be integrated with additional row merge rules in place. 
\end{itemize}

\item  Shoe Track Key Table
    
    \begin{itemize}
    \item Most of this table is not used for the MASP calculation, and the only row of data comes from source B. 
\end{itemize}

\item  Well Data Key Table
    
    \begin{itemize}
    \item There is only one row in this table because the data is merged based on the well name. This ensures both source A and source B are performing the MASP calculation for the same well.
\end{itemize}

\end{itemize}

\subsection{Discussion}

Our main result from the case study is that two engineers were able to conduct the same engineering analysis, without ever communicating with each other, and compose their individual models to create an integrated spreadsheet with integrated perspective. Specifically, two engineers performed a MASP calculation, which is a key aspect of well construction and is performed on every well around the world in a variety of fashions. The integrated result guarantees that individual ologs and integrated ologs are compliant with components of the MASP calculation technical requirements, without pre-coordination and consensus.  As a bonus, we also obtain a semantics-preserving data-exchange between the source sheets. 

Engineering teams largely go about integration and data exchange today by treating subject matter experts as the personification of requirements, which are originally written in natural language as text inside documents that we call global standards or regulations. The current mainstream approach taken by industry is to try to organize these requirements by extracting them from documents and placing them into more sophisticated management systems (e.g., a database where each requirement is an object, and a user interface that includes workflows to create, edit, and approve requirements). This approach relies on the idea that after the requirements are organized in the requirement management system, engineers will be able to convert them into rules that can be incorporated into computable models. The challenge to do this, so far, has been the lack of an adequate language that is expressive enough to capture both the general nature of an engineering requirement and its specific computational nature.

By representing objects and models as ologs, engineers can define data structure in a way that makes sense to them. Those structures can then be translated to a shared space within which these objects reside, using automated methods for their construction and inter-relation.  We pick up this line of thought in the conclusion.

\subsection{Governance}

Left as open questions are the authorship and/or governance of the overlap ologs required to integrate  sheets-- either of the original engineers can create (potentially many, depending on the desired integration semantics) such overlaps, as can 3rd parties (and in fact, CQL can automatically suggest overlap between ologs).  In fact, at a formal level, managing an overlap between two ologs is ``dual'' to managing a collection of WHERE clauses in SQL queries that join two databases together (the algebraic integration semantics of Figure~\ref{fig:algint} being formally dual to SQL's join semantics as well).  As such, governance of overlaps is too large of a topic to go into here. Regardless of authorship, semantics-preservation is guaranteed without requiring consensus by relying on the formulas in the sheets to fully define semantics.

    \subsection{Generalization to Multi-Model Merge}    
    
    In this paper we have focused on an olog merge of the form $\cdot \leftarrow \cdot \rightarrow \cdot$, where the center dot is the overlap olog and the other two dots are the source olog.  However, all of the theory and algorithms in this paper generalize to diagrams of any shape
    \cite{schultz_wisnesky_2017}, such as
    $$
\xymatrix{
& \cdot \ar[ld] \ar[rd] & & \cdot \ar[ll] \ar[rr] & & \cdot \ar[ld] \ar[rd] & \\
\cdot & & \cdot & & \cdot & & \cdot }    
   $$
It is easy to see that a data integration of the above form requires only logarithmically many mappings in the number of nodes, rather than quadratically many that would be required if all nodes were to be mapped to all nodes.  It is because our methodology can merge any shape of olog mappings, such as the above, that we say it does not require every source to be related to every source.

\section{Generating and Checking Verification Conditions}
\label{sec:vc}

This section will be of interest primarily to computer scientists.

\subsection{Functoriality Conditions}

In this section we describe how to generate and solve the verification conditions required to ensure the semantic consistency of the source schemas with respect to the overlap schema.  The verification conditions associated to schema  mapping $M_A : S_O \to S_A$ ensure/are that $S_A$ proves $M_A(p) = M_A(q)$ for every equation $p = q$ of $S_O$, and similarly for $M_B$.  This establishes that $M_A$ and $M_B$ are functors and hence that we can proceed with algebraic data integration.  In Figure~\ref{fig:coq} we display these verification conditions for $M_A$ and $M_B$ in the Coq proof assistant~\cite{citeulike:14251820} where they can be proved by humans.  The CQL tool also emits the verification conditions in TPTP form~\cite{tptp} so that they can be input to a variety of automated theorem provers~\cite{10.5555/280474}.  It took one of the authors approximately four hours to prove all the generated conditions in this paper in Coq with elementary methods, using a handful of basic assumptions about e.g. commutativity of addition, but with no domain expertise.  Although checking functoriality is undecidable in general, the conditions can always be decided for a particular model, providing a slightly weaker but still useful guarantee. 


\begin{figure}[p]
    \centering
    \begin{footnotesize}
    \begin{verbatim}
//Schema A    
Parameters String Float : Type.
Parameters plus times : Float * Float -> Float.
...
Parameters CasingSection ZoneofInterest MASPCalcStep1 : Set.
...
Parameter HeaderInfo_TypeofWell : HeaderInfo -> String.
Parameter HeaderInfo_RigName : HeaderInfo -> String.
Parameter HeaderInfo_Field : HeaderInfo -> String.
...
Axiom ax0 : forall (x : HeaderInfo),  HeaderInfo_RKBML(x) = plus(HeaderInfo_WaterDepth(x), HeaderInfo_RKBHeight(x)).
Axiom ax1 : forall (x : MASPCalcStep2a),  MASPCalcStep2a_MASPShoe(x) =
    minus(minus(minus(MASPCalcStep2a_FracPressureatDeepestShoe(x), MASPCalcStep2a_MudHydrostaticShoe(x)), 
        MASPCalcStep2a_GasHydrostaticShoe(x)), MASPCalcStep2a_SWHydrostatic(x)).
Axiom ax2 : forall (x : MASPCalcStep2a),  MASPCalcStep2a_InterfaceTVDBHP(x) =
    plus(HeaderInfo_RKBML(MASPCalcStep2a_Well(x)),
        times(MASPCalcStep2a_GasRatio(x), minus(FGPPInputs_RKBTVD(MASPCalcStep2a_TVDDeepestOH(x)),
        HeaderInfo_RKBML(MASPCalcStep2a_Well(x))))).
Axiom ax3 : forall (x : MASPCalcStep2b),  MASPCalcStep2b_MAWP(x) =
    plus(MASPCalcStep2a_MinimumMASP(MASPCalcStep2b_ReferenceMASP(x)),
    MASPCalcStep2a_SWHydrostatic(MASPCalcStep2b_ReferenceMASP(x))).
...
Definition t_Step1_pb : MASPCalcStep1 -> Float := fun x => CasingSection_BurstRating(MASPCalcStep1_CasingSection(x)).
Definition t_Step1_pmud : MASPCalcStep1 -> Float := fun x => IntervalInfo_DownholeMudWeight(MASPCalcStep1_Interval(x)).
Definition t_Step1_ppore : MASPCalcStep1 -> Float := fun x => ZoneofInterest_BackupPorePressure(MASPCalcStep1_ZoneName(x)).
...
Conjecture conj22 : forall (i : MASPCalcStep1),  t_Step1_casingburst(i) =
    minus(times(t_Step1_seventy(i), t_Step1_pb(i)), 
    times(t_Step1_tvd(i), times(t_Step1_pointohfivetwo(i),
    minus(t_Step1_pmud(i), t_Step1_ppore(i))))).
Conjecture conj23 : forall (i : MASPCalcStep2a),  t_Step2_maspbhp(i) =
    minus(t_Step2_bhp(i), plus(t_Step2_mhs(i), plus(t_Step2_ghs(i), t_Step2_swhs(i)))).
...
\end{verbatim}
\hrulefill
\begin{verbatim}
//Schema B
Parameters String Float : Type.
Parameters plus times : Float * Float -> Float.
...
Parameters WellDataKey CasingKey BurstCalculationKey MASPShoeKey MASPKey : Set
...
Parameter WellDataKey_Block : WellDataKey -> String.
Parameter WellDataKey_Well : WellDataKey -> String.
Parameter WellDataKey_WellBore : WellDataKey -> String.
...
Parameter CasingKey_ItemofInterest : CasingKey -> ItemofInteerestKey.
Parameter CasingKey_PPFPKey : CasingKey -> PorePressureFracPressureKey.
...
Axiom ax0 : forall (x : CasingKey),  CasingKey_MaxPorePressureatHoleSectionTDppg(x) =
    PorePressureFracPressureKey_PorePressuremidppg(CasingKey_PPFPKey(x)).
Axiom ax1 : forall (x : BurstCalculationKey),  BurstCalculationKey_TVD(x) =
    ItemofInteerestKey_ItemofInterestDepthRKBTVDFt(BurstCalculationKey_ItemofInterest(x))
Axiom ax2 : forall (x : MASPShoeKey),  MASPShoeKey_FracGradientatDeepestExposedShoeppg(x) =
    ExposedShoeKey_FracGradientatDeepestExposedShoeppg(MASPShoeKey_ExposedShoeKey(x)).
Axiom ax3 : forall (x : MASPKey),  MASPKey_MASPOpenhole(x) =
    MASPOpenHoleKey_MASPOpenHole(MASPKey_MASPOpenHoleKey(x)).
...
Definition t_Step1_pb : BurstCalculationKey -> Float := 
    fun x => ItemofInteerestKey_ItemofInterestBurstRatingpsi(BurstCalculationKey_ItemofInterest(x)).
Definition t_Step1_pmud : BurstCalculationKey -> Float := fun x => BurstCalculationKey_DHEMW(x).
Definition t_Step1_ppore : BurstCalculationKey -> Float := fun x => BurstCalculationKey_BacksideEMW(x).
...
Conjecture conj21 : forall (i : MASPKey),  t_Step2b_maspshoe(i) = 
    minus(t_Step2b_bhp(i), plus(t_Step2b_mhs(i), plus(t_Step2b_ghs(i), t_Step2b_swhs(i)))).
Conjecture conj22 : forall (i : BurstCalculationKey),  t_Step1_casingburst(i) =
    minus(times(t_Step1_seventy(i), t_Step1_pb(i)), 
    times(t_Step1_tvd(i), times(t_Step1_pointohfivetwo(i), minus(t_Step1_pmud(i), t_Step1_ppore(i))))).
Conjecture conj23 : forall (i : MASPKey),  t_Step2_maspbhp(i) = 
    minus(t_Step2_bhp(i), plus(t_Step2_mhs(i), plus(t_Step2_ghs(i), t_Step2_swhs(i)))).
\end{verbatim}
\end{footnotesize}
    \caption{Coq Verification Conditions for Mapping Functoriality}
    \label{fig:coq}
\end{figure}

\subsection{Conservativity/Consistency Conditions}

In this section we describe how to generate and solve the verification conditions required to ensure the semantic consistency of the integrated result olog.  The integrated result is a quotient of a coproduct of input databases; to compute the quotient, a step-by-step process is used, where at each step a new olog is constructed from the previous one (starting from the coproduct of the inputs and terminating on the integrated result).  At each step of the process an equational theory representing the result is modified by adding additional equations and/or symbols~\cite{schultz_wisnesky_2017}.  For example, at one step we might add $Alice.age = 20$ and at another step add $Bob.age = 30$ and at another step add $Alice = Bob$, from which it follows that $20=30$-- a contradiction, or rather, a ``non-conservative extension'' of our original definition of numbers, for which $20$ and $30$ are not equal (number systems in which $20=30$ are found in e.g. modular arithmetic and can in fact be represented by ologs).  At every step of the integration computation, CQL checks that there are no equations that reduce to $c_1 = c_2$ for distinct numerals, strings, etc under the usual reduction rules for arithmetic such as $1+1 \mapsto 2$ etc, providing significant defense against contradictions.  In general, contradiction detection is undecidable but definitional/free spreadsheets (see section~\ref{sec:ologs}) are always contradiction-free and CQL can check ologs for freeness, providing more protection still. 

\section{Conclusion and Applications}
\label{sec:conc}

Our case study is an example of a non-human-consensus-based method for merging engineering models, built on the idea that formality drives consensus in the semantic sense even if the people involved never meet.  We conclude by describing a particular enterprise example use case and then describe related work. 

\subsection{An Example Use Case}

Our use case is to certify that an (equational) engineering model complies with (equational) requirements, a role today performed by human auditors working with English text and/or spreadsheets.  We focus on this use case because it is the smallest use case that we can think of that is captured by any part of this paper's case study, but as we will see it can still have significant impact.  In particular, if we have $R$ (equational) requirements to certify on $W$ (equational) models each year at a cost of $C$ per certification then the direct cost savings of this use case is $T\times W \times C$ annually, a number that can immediately be seen to be large in large enterprises when there are hundreds of thousands of (equational) requirements to check each year.  

To implement the use case we require two olog schemas, $S$ and $T$, and a schema mapping $F : S \to T$.  For example, we may let $S$ be this paper's overlap schema $S_O$ (Figure~\ref{fig:schemaO}) and let $T$ be this paper's source schema A, namely, $S_A$ (Figure~\ref{fig:schemaA}) (or source schema B, namely, $S_B$, Figure~\ref{fig:schemaB}) and let $F$ be this paper's schema mapping $S_O \to S_A$, shown in Figure~\ref{fig:mapA} (resp. Figure~\ref{fig:mapB} for source B).  The core of the use case is to check if $F$ denotes a functor, i.e., check if $F$ translates the equations of $S$ into $T$ in a way that preserves provability as described in Section~\ref{sec:vc}.  If so, then we say that $T$ complies with the requirements of $S$ as viewed by $F$.  Optionally, we may additionally take as input an $S$-Instance $X$ and $T$-Instance $Y$ and check that $X = F ; Y$, which in this case study corresponds to checking that the equations satisfied by the overlap instance $I_O$ hold in the source instance $I_A$ (resp.  in $I_B$) once translated along $F$.  That is, our use case can apply to models without data (equations only) as well as to models with data.

The discussion in the preceding paragraph does not describe any method for constructing the inputs to the use case.  One option is to simply build up a library of ologs and olog mappings manually, and with composition, such as we did in this paper.  Another option is to heuristically generate ologs and mappings from spreadsheets, or even English-text, on-the-fly, to be reviewed/modified by a human-- in general, choice will be required among various possible oligifications of a given spreadsheet and among the various possible mappings between ologs.  Such heuristics are the subject of future work, but no matter which ologs and mappings are chosen, we can be confident that if the functoriality check succeeds then the (equational) requirements are met.  And we are confident that there are many more use cases besides this one waiting to be discovered.

\subsection{Related Work}

Besides relations to algebraic databases and traditional data integration~\cite{schultz_wisnesky_2017}, our algorithm is related to the idea of extending spreadsheets with deductive logic, for example with prolog, an idea whose long history is described in~\cite{excelrel}.  Indeed, our rules in this paper are expressible in prolog, at least when prolog is understood to include excel's arithmetic operations, which often it isn't (it is unclear how much arithmetic is modeled in~\cite{excelrel} fop example).  However, unlike any work in~\cite{excelrel}, our goal is to connect multiple spreadsheets, not query different parts of the same sheet, which constrains the form of the rules we can write to a subset of prolog, and also constrains the vocabulary we can use within prolog.  That is, the algebraic design pattern of Figure~\ref{fig:algint} says which rules to write, whereas~\ref{fig:algint} provides no guidance.

{\bf Acknowledgements}. The authors would like to thank Chevron's Ben Randell for the support and leadership to conduct this case study, and also Eswaran (Sub) Subrahmanian, Spencer Breiner, and Priyaa Srinivasan of NIST for their help on understanding the relationship between category theory and compositional structures for systems engineering and design.  The authors would like to thank Conexus's Joshua Meyers for help implementing the paper described in this software.



\newpage

\section{Appendix: Category theory}
\label{sec:cat}

In this section, we review standard definitions and results from category theory~\cite{MR2229319}.  A {\it quiver}, (aka directed multi-graph) $Q$ consists of
 a class ${\sf Ob}(Q)$, the members of which we call {\it objects} (or {\it nodes}), and
for all objects $c_1, c_2$, a set $Q(c_1, c_2)$, the members of which we call {\it morphisms} (or {\it arrows}) from $c_1$ to $c_2$.  We may write $f : c_1 \to c_2$ or $c_1 \to_f c_2$ instead of $f \in C(c_1,c_2)$.   For an arrow $f:c_1\to c_2$ in a quiver, we call
  $c_1$ the {\it source} of $f$ and $c_2$ the {\it target} of
  $f$.  In a quiver $Q$, a {\it path} from $c_1$ to $c_k$
  is a non-empty finite list of nodes and arrows
  $c_1\To{e_1}c_2\To{e_2}\cdots\To{e_{k-1}}c_k$. A {\it category} $C$ is a quiver equipped with the
  following structure:
  \begin{itemize}
\item for all objects $c_1,c_2,c_3$, a function $\circ_{c_1,c_2,c_3} : C(c_2,c_3) \times C(c_1,c_2) \to C(c_1,c_3)$, which we call {\it composition}, and for which we write $x ; y$ to mean $y \circ x$, and
\item for every object $c$, an arrow ${\sf id}_c \in C(c,c)$, which we call the {\it identity} for $c$. 
\end{itemize}
We may drop subscripts on ${\sf id}$ and $\circ$, when doing so does not create ambiguity.  These data  must obey axioms stating that $\circ$ is associative and ${\sf id}$ is its unit:
$$
{\sf id} \circ f = f \ \ \ \ f \circ {\sf id} = f \ \ \ \ f \circ (g \circ h) = (f \circ g) \circ h
$$

An object $c$ of a category $C$ is called {\it initial} if for all $c'\in C$, there is a unique morphism $c\to c'$. Dually, an object $c$ of a category is called {\it final}/{\it terminal} if for all $c'\in C$, there is a unique morphism $c'\to c$. A {\it functor} $F : C \to D$ between categories $C$ and $D$ consists of:
\begin{itemize}
    \item a class function $F : {\sf Ob}(C) \to {\sf Ob}(D)$, and
    \item for every $c_1,c_2 \in {\sf Ob}(C)$, a function $F_{c_1,c_2} : C(c_1,c_2) \to D(F(c_1),F(c_2))$, where we may omit object subscripts when they can be inferred, such that 
\end{itemize}
$$
F({\sf id}_c) = {\sf id}_{F(c)} \ \ \ \ \ \ F(f \circ g) = F(f) \circ F(g).
$$

A {\it natural transformation} $h : F \to G$ between functors $F, G : C \to D$ consists of a family of morphisms $h_c : F(c) \to G(c)$, indexed by objects in $C$, called the {\it components} of $h$,
 such that for every $f : c_1 \to c_2$ in $C$ we have
$
h_{c_2} \circ F(f) = G(f) \circ h_{c_1}
$.  The family of equations defining a natural transformation may be depicted as a {\it commutative diagram}: 
$$
\xymatrix{
F(c_1) \ar[d]_{h_{c_1}} \ar[r]^{F(f)} & F(c_2) \ar[d]^{h_{c_2}} \\
G(c_1) \ar[r]_{G(f)}& G(c_2) \\ 
}
$$
The commutativity of such a diagram means that any two parallel
paths in the diagram have the same composition in $D$.  A {\it pushout} of objects $A,B,C$ and morphisms $f,g$ in a category, as shown below, is an object $D$ and morphisms $\alpha$ and $\beta$ as shown below, having the universal property that for any other such $D'$ and $\alpha'$ and $\beta'$, there is a unique morphism $\theta$ making the diagram commute:
\[
\xymatrix{
 A \ar[r]^{g}      \ar[d]_{f}         & C \ar[d]_{\beta} \ar@/^1pc/[rdd]^{\beta'}   \\
 B \ar[r]^{\alpha} \ar@/_1pc/[rrd]_{\alpha'} & D \ar@{-->}[rd]^{\theta} \pullbackcorner[ul] \\
 && D' 
}
\]

The dual notion of pushout is {\it pullback}.  A pushout over an initial object is called a co-product.  A pullback over a final object is called a product.  Pushouts generalize to arbitrary diagrams, in which case they are called colimits, but we do not define colimits here.   Given two functors $F : C \to D$ and $G : D \to C$, we say that $F$
  is {\it left adjoint} to $G$, written $F \dashv G$, when for every
  object $c$ in $C$ and $d$ in $D$ that the set of morphisms
  $F(c) \to d$ in $D$ is isomorphic to the set of morphisms
  $c \to G(d)$ in $C$, naturally in $c$ and $d$ (i.e., when we
  independently consider each side of the isomorphism as a functor
  $C \to {\sf Set}$ and as a functor ${\sf D} \to {\sf Set}$).
  
\section{Appendix: Equational Logic and Universal Algebra}
\label{sec:alg}

In this section we review standard material on multi-sorted equational logic, following the exposition in \cite{schultz_wisnesky_2017}.  A {\it signature} $Sig$ consists of:
\begin{enumerate}
\item A set $Sorts$ whose elements are called {\it sorts},
\item A set $Symbols$ of pairs $(f, s_1 \times \ldots \times s_k \to s)$ with $s_1, \ldots, s_k , s \in Sorts$ and no $f$ occurring in two distinct pairs.  We write $f \taking X$ instead of $(f, X) \in Symbols$.  When $k = 0$, we may call $f$ a {\it constant symbol} and write $f \taking s$ instead of $ f : \ \to s$.  Otherwise, we may call $f$ a {\it function symbol}.  
\end{enumerate}
We assume we have some countably infinite set $\{ v_1, v_2, \dots \}$, whose elements we call {\it variables} and which are assumed to be distinct from any sort or symbol we ever consider.  A {\it context} $\Gamma$ is defined as a finite set of variable-sort pairs, with no variable given more than one sort:
$$
\Gamma := \{ v_1 : s_1, \ldots, v_k : s_k \}
$$
When the sorts $s_1 , \ldots, s_k$ can be inferred, we may write a context as $\{ v_1, \ldots, v_k \}$.  We may write $\{ v_1 : s, \ldots, v_k : s \}$ as $\{ v_1, \ldots, v_k : s\}$.  We may write $\Gamma \cup \{ v : s \}$ as $\Gamma, v:s$.  We inductively define the set $Terms^s(Sig, \Gamma)$ of {\it terms} of sort $s$ over signature $Sig$ and context $\Gamma$ as:
\begin{enumerate}
\item $x \in Terms^s(Sig, \Gamma)$, if $x : s \in \Gamma$, 
\item $f(t_1, \ldots, t_k)  \in Terms^s(Sig, \Gamma)$, if $f \taking s_1 \times \ldots \times s_k \to s$ and $t_i \in Terms^{s_i}(Sig, \Gamma)$ for $i = 1, \ldots , k$. When $k=0$, we may write $f$ for $f()$.  When $k=1$, we may write $t_1.f$ instead of $f(t_1)$.  When $k=2$, we may write $t_1 \ f \ t_2$ instead of $f(t_1,t_2)$.
\end{enumerate}
We refer to $Terms^s(Sig, \emptyset)$ as the set of {\it ground} terms of sort $s$.  We will write $Terms(Sig, \Gamma)$ for the set of all terms in context $\Gamma$, i.e., $\bigcup_s Terms^s(Sig, \Gamma)$.  An {\it equation} over $Sig$ is a formula $\forall \Gamma. \ t_1 = t_2 : s$ with $t_1, t_2 \in Terms^s(Sig,\Gamma)$; we will omit the $: s$ when doing so will not lead to confusion.  A {\it theory} $Th$ is a pair of a signature and a set of equations over that signature.   Associated with a theory $Th$ is a binary relation between (not necessarily ground) terms, called {\it provable equality}.  We write $Th \vdash \forall \Gamma. \ t = t' : s$ to indicate that the theory $Th$ proves that terms $t, t' \in Terms^s(Sig, \Gamma)$ are equal according to the usual rules of multi-sorted equational logic.  From these rules it follows that provable equality is the smallest equivalence relation on terms that is a congruence, closed under substitution, closed under adding variables to contexts, and contains the equations of $Th$.  A {\it morphism of signatures} $F \taking Sig_1 \to Sig_2$ consists of:
\begin{itemize}
\item a function $F$ from sorts in $Sig_1$ to sorts in $Sig_2$, and
\item a function $F$ from function symbols $f \taking s_1 \times \ldots \times s_n \to s$ in $Sig_1$ to terms in 
$$Terms^{F(s)}(Sig_2, \{ v_1 \taking F(s_1), \ldots, v_n \taking F(s_n) \} ).$$
To clearly indicate the context $\{ v_1, \ldots, v_n \} $, the function $F(f)$ may be written in ``$\lambda$ notation'', i.e. as $F(f) = \lambda v_1, \ldots, v_n . g(v_1, \ldots, v_n)$ for some term $g$, where the $\lambda$ is omitted if $n=0$.
\end{itemize}

For example, let $Sig_1$ consist of two sorts, $a, b$, and one function symbol, $f \taking a \to b$, and let $Sig_2$ consist of one sort, $c$, and one function symbol, $g\taking c \to c$.  There are countably infinitely many morphisms $F \taking Sig_1 \to Sig_2$, one of which is defined as $F(a) := c$, $F(b) := c$, and $F(f) := \lambda v \taking c . \ g(g(v))$.  In the literature on algebraic specification, our definition of signature morphism is called a ``derived signature morphism''~\cite{conf/wadt/MossakowskiKM14}.

A {\it morphism of theories} $F : Th_1 \to Th_2$ is a morphism of signatures that preserves provability:
$$
Th_1 \vdash \forall v_1 : s_1, \ldots v_n : s_n. \ t_1 = t_2 : s \ \ \ \Rightarrow \ \ \ Th_2 \vdash \forall v_1 : F(s_1), \ldots, v_n : F(s_n) . \ F(t_1) = F(t_2) : F(s)
$$
where we have extended $F$ to operate on terms.  An {\it algebra} $A$ over a signature $Sig$ consists of:
\begin{itemize}
\item a set of {\it carriers} $A(s)$ for each sort $s$, and
\item a function $A(f) : A(s_1) \times \ldots \times A(s_k) \to A(s)$ for each symbol $f : s_1 \times \ldots s_k \to s$.
\end{itemize}
Let $\Gamma := \{ v_1 : s_1 , \ldots , v_n : s_n \}$ be a context.  An $A$-{\it environment} $\eta$ for $\Gamma$ associates each $v_i$ with an element of $A(s_i)$.  The meaning of a term in $Terms(Sig, \Gamma)$ relative to $A$-environment $\eta$ for $\Gamma$ is recursively defined as:
$$
A \llbracket v\rrbracket \eta = \eta(v) \ \ \ \ \ \ \ A \llbracket f(t_1, \ldots, t_n) \rrbracket \eta = A(f)(A\llbracket t_1 \rrbracket \eta , \ldots, A \llbracket t_i \rrbracket \eta)
$$
An algebra $A$ over a signature $Sig$ is a {\it model} of a theory $Th$ on $Sig$ when $Th \vdash \forall \Gamma. \ t = t' : s$ implies $A\llbracket t \rrbracket\eta = A\llbracket t' \rrbracket\eta$ for all terms $t, t' \in Terms^s(Sig, \Gamma)$ and $A$-environments $\eta$ for $\Gamma$.  Deduction in multi-sorted equational logic is sound and complete: two terms $t, t'$ are provably equal in a theory $Th$ if and only if $t$ and $t'$ denote the same element in every model of $Th$.  

From a signature $Sig$ we form its {\it term algebra} $\llbracket Sig \rrbracket$ as follows.  The carrier set $\llbracket Sig \rrbracket(s)$ is defined as the set of ground terms of sort $s$.  The function $\llbracket Sig \rrbracket(f)$ for $f : s_1 \times \ldots s_k \to s$ is defined as the function $t_1 , \ldots t_n \mapsto f(t_1, \ldots, t_n)$.  From a theory $Th$ on $Sig$ we define its {\it term model} $\llbracket Th \rrbracket$ to be the quotient of $\llbracket Sig \rrbracket$ by the equivalence relation $Th \vdash$.  In other words, the carrier set $\llbracket Th \rrbracket(s)$ is defined as the set of equivalence classes of ground terms of sort $s$ that are provably equal under $Th$.  The function $\llbracket Th \rrbracket(f)$ is $\llbracket Sig \rrbracket (f)$ lifted to operate on equivalence classes of terms.  To represent $\llbracket Th \rrbracket$ on a computer, or to write down $\llbracket Th \rrbracket$ succinctly, we must choose a {\it representative} for each equivalence class of terms; this detail can be ignored in this paper.

A {\it morphism of algebras} $h : A \to B$ on a signature $Sig$ is a family of functions $h(s) : A(s) \to B(s)$ indexed by sorts $s$ such that:
$$
h(s)(A(f)(a_1, \ldots, a_n)) = B(f)(h(s_1)(a_1), \ldots, h(s_n)(a_n)) 
$$
for every symbol $f : s_1 \times \ldots \times s_n \to s$ and $a_i \in A(s_i)$.  We may abbreviate $h(s)(a)$ as $h(a)$ when $s$ can be inferred.  The term algebras for a signature $Sig$ are initial among all $Sig$-algebras: there is a unique morphism from the term algebra to any other $Sig$-algebra.  Likewise,  term models are initial among models. 

\section{Appendix: The Categorical Theory of Algebraic Databases}
\label{sec:cql} 
In this section we briefly describe the categorical theory of algebraic databases, following ~\cite{schultz_wisnesky_2017}. We first fix a theory, $\mathit{Ty}$, called the {\it type side} of our formalism. The sorts of $\mathit{Ty}$ are called {\it types} and the functions of $\mathit{Ty}$ are the functions that can appear in schemas and instances.  The intended meaning of this theory, $\llbracket Ty \rrbracket$, is a category with products.  A {\it schema} $S$ on type side $\mathit{Ty}$ is a theory extending $\mathit{Ty}$ with new sorts (called {\it entities}), new unary functions from entities to types (called {\it attributes}), new unary functions from entities to entities (called {\it foreign keys}), and new equations (called {\it data integrity constraints}) of the form $\forall v:s. \ t = t'$, where $s$ is an entity and $t,t'$ are terms of the same type, each containing a single free variable $v$.  The intended meaning of this theory, $\llbracket S \rrbracket$, is a category extending $\llbracket Ty \rrbracket$.  Let $S$ and $T$ be schemas on the same type side $\mathit{Ty}$.  A schema mapping $F : S \to T$ is defined as a ``derived signature morphism''~\cite{conf/wadt/MossakowskiKM14} from $S$ to $T$ that is the identity on $\mathit{Ty}$.  That is, $F : S \to T$ assigns to each entity $e \in S$ an entity $F(e) \in T$, and to each attribute / foreign key $f : s \to s'$ a term $F(f)$, of type $F(s')$ and with one free variable of type $F(s)$, in a way that respects  equality: if $S \vdash t = t'$, then $T \vdash F(t) = F(t')$.  The intended meaning of $F$ is a functor $\llbracket S \rrbracket \to \llbracket T \rrbracket$ that is the identity on $Ty$.  The collection of all schema mappings forms a category with colimits.  

An {\it instance} $I$ on schema $S$ is a theory extending $S$ with new 0-ary function (constant) symbols called {\it generators} and non-quantified equations.  The intended meaning of an instance $I$, written $\llbracket I \rrbracket$, is the {\it term model} (i.e., {\it initial algebra}) for $I$ which contains, for each sort $s$, a {\it carrier set} consisting of the closed terms of sort $s$ modulo provability in $I$.  That is, the intended meaning of an $S$-instance $I$ is a functor $\llbracket S \rrbracket \to Set$, namely, the initial such functor in the category of all functors consistent with $I$.  Let $I$ and $J$ be instances on the same schema $S$.  A data mapping $h : I \Rightarrow J$ is defined as a ``derived signature morphism''~\cite{conf/wadt/MossakowskiKM14} from $I$ to $J$ that is the identity on $\mathit{S}$.  That is, $h : I \Rightarrow J$ assigns to each generator $g : s$ in $I$ a closed term $h(g) : s$ in $J$ in a way that respects equality: if $I \vdash t = t'$, then $J \vdash h(t) = h(t')$.  A morphism of instances thus denotes homomorphism (natural transformation) of algebras $\llbracket I \rrbracket \to \llbracket J \rrbracket$.  The database instances and morphisms on a schema $S$ constitute a category with all colimits, denoted $S\iinst$, and a schema mapping $F : S \to T$ induces a functor $\Sigma_F : S\iinst \to T\iinst$ defined by substitution.  The functor $\Sigma_F$ has a right adjoint, $\Delta_F : T\iinst \to S\iinst$, which corresponds to composition when we are thinking semantically: $\llbracket \Delta_F(I) \rrbracket = \llbracket F \rrbracket ; \llbracket I \rrbracket$.  Semantically, $\llbracket \Sigma_F(I) \rrbracket$ computes the ``left Kan-extension'' of $\llbracket I \rrbracket$ along $\llbracket F \rrbracket$~\cite{schultz_wisnesky_2017}.  

\newpage

 \begin{figure}[p]

 \includegraphics[width=6.5in]{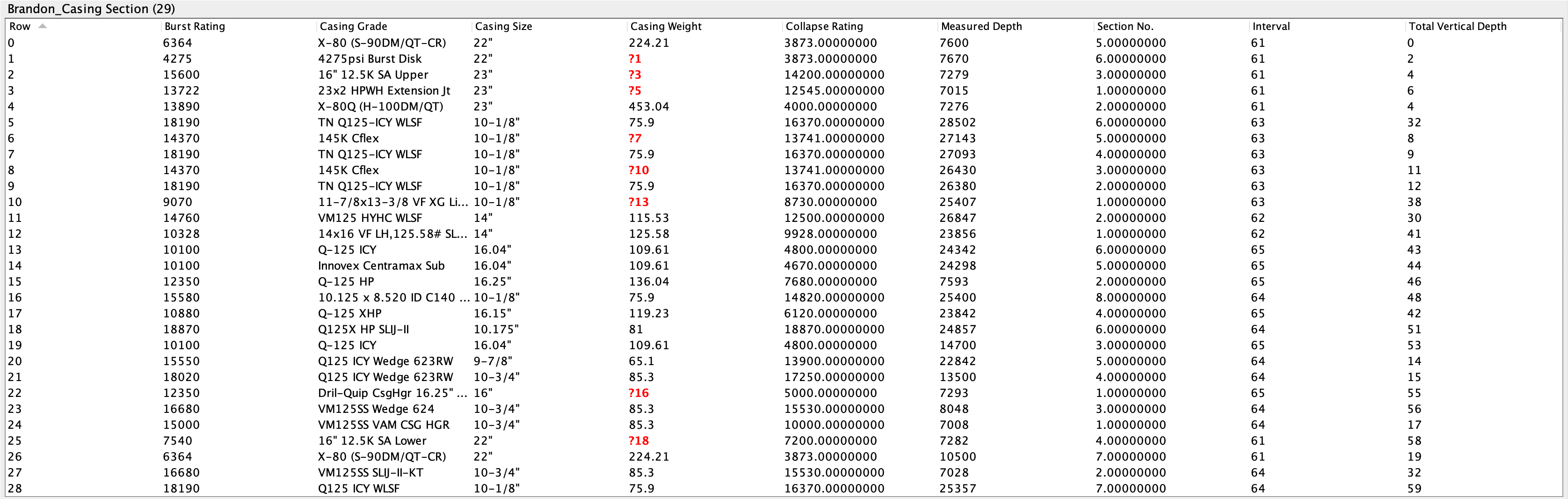} 
 
\includegraphics[width=6.5in]{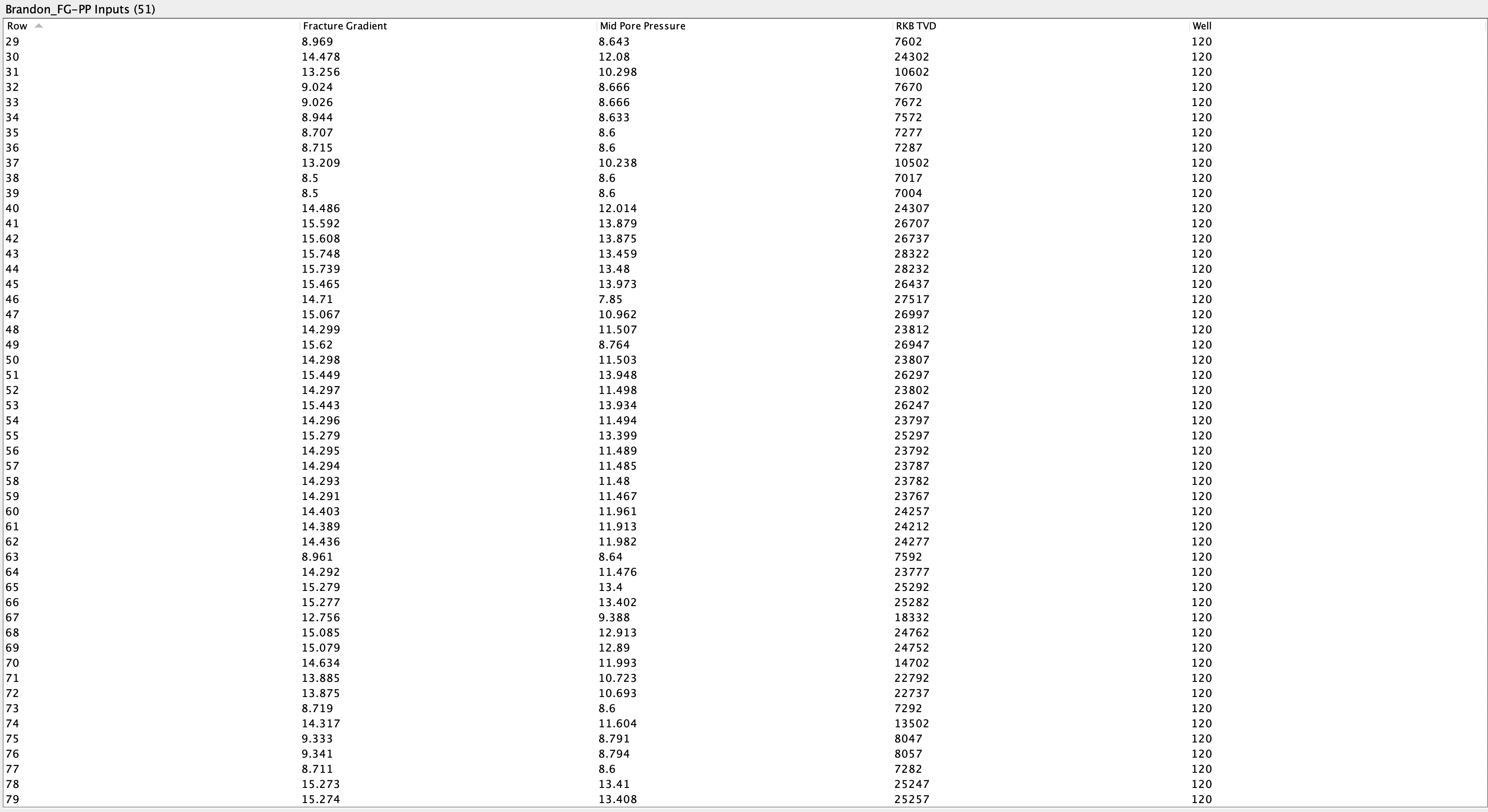}

\caption{Full Integrated Olog, 1 of 4}
\label{fig:intA}
\end{figure}
\newpage    

     \begin{figure}[p]
\includegraphics[width=6.5in]{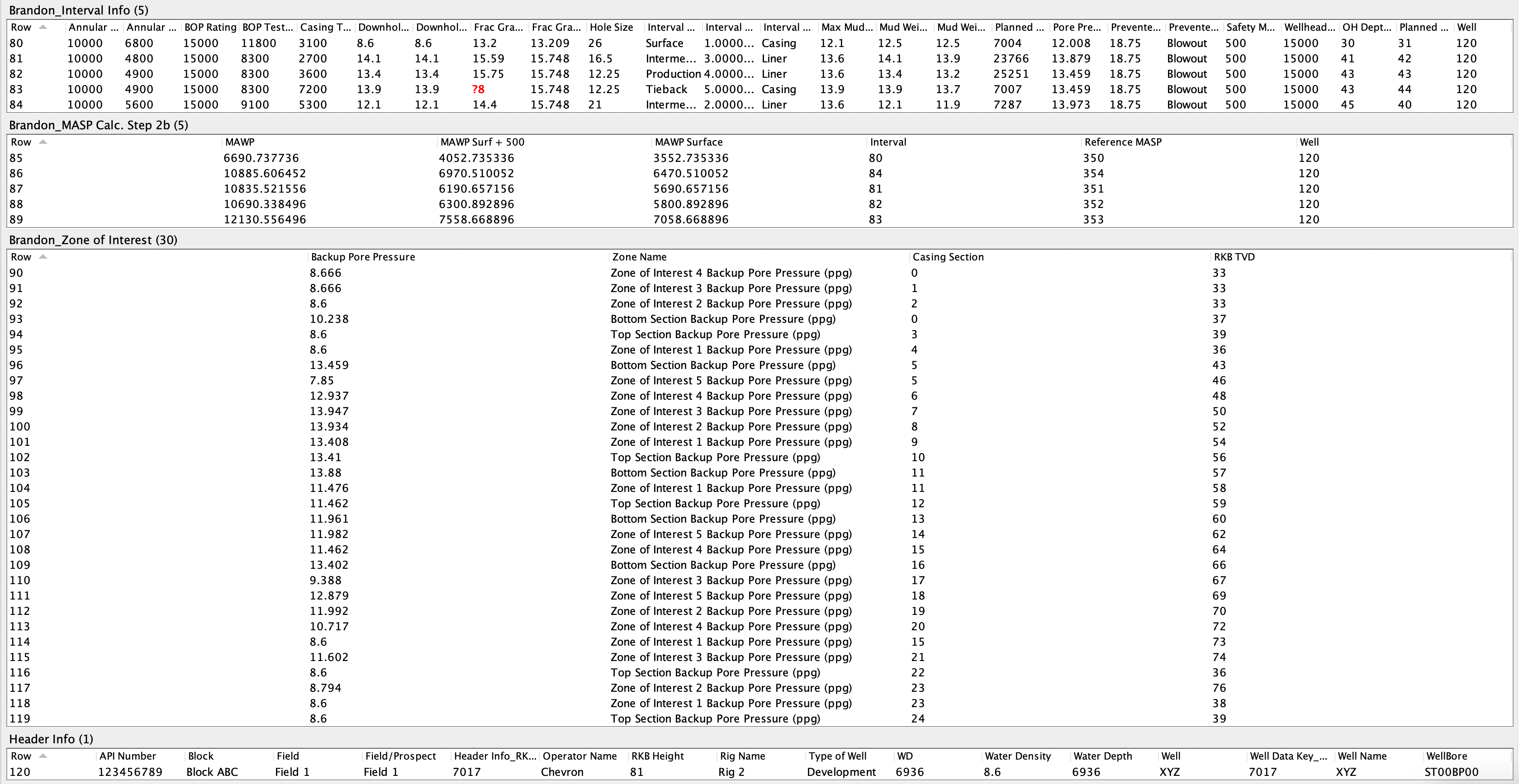}
\includegraphics[width=6.5in]{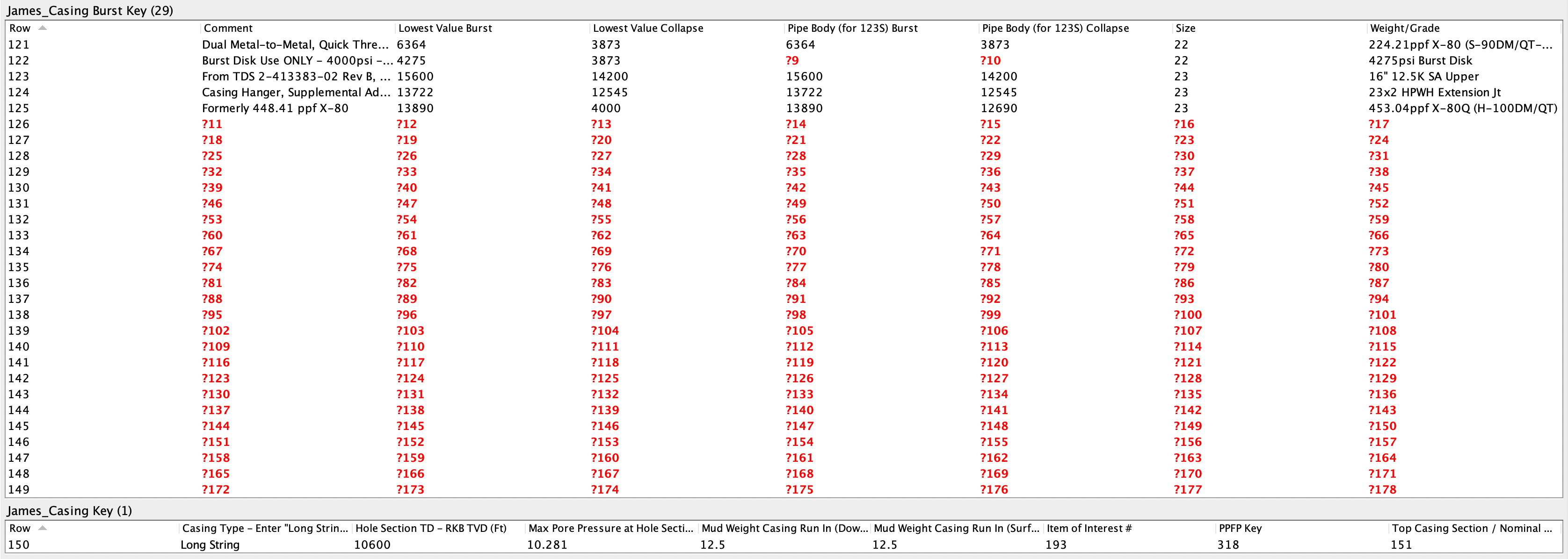}
\includegraphics[width=6.5in]{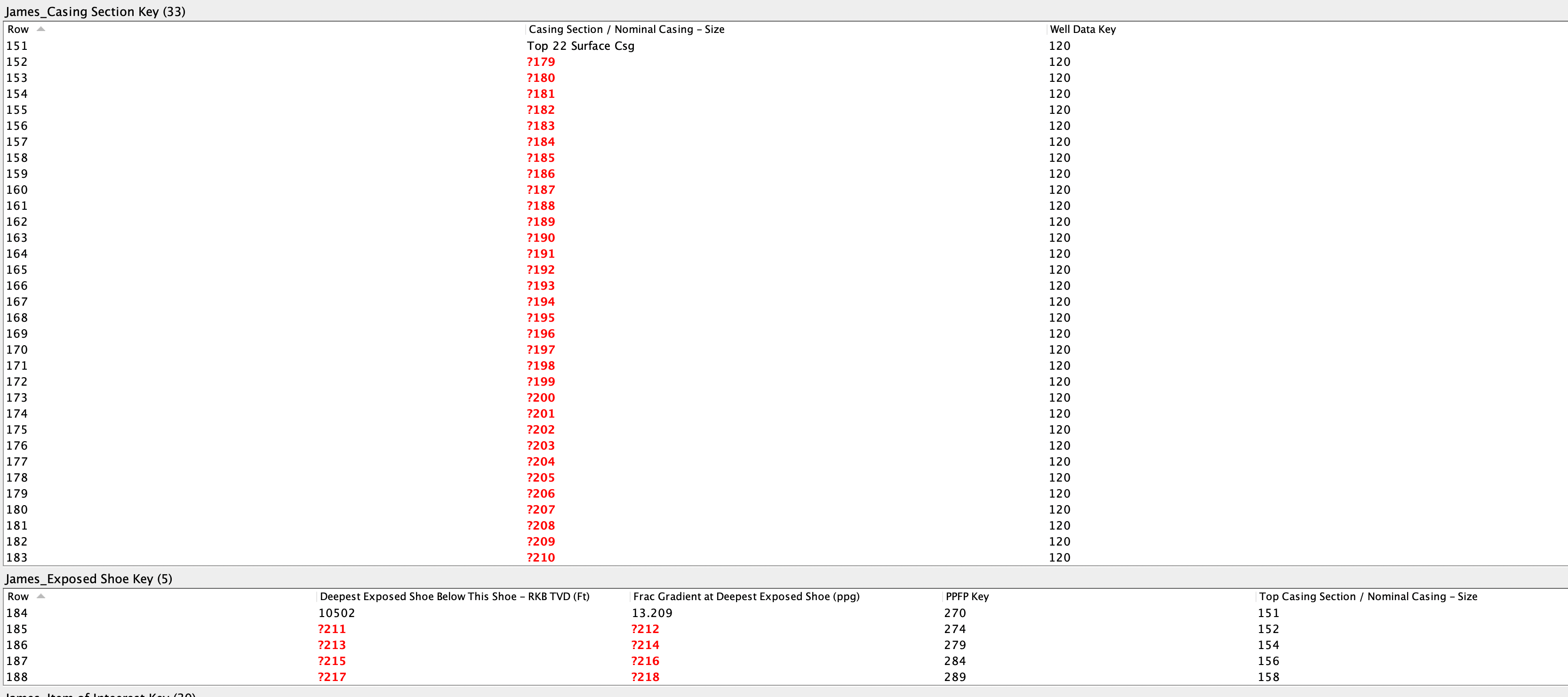}

        \caption{Full Integrated Olog, 2 of 4}
        \label{fig:intB}
    \end{figure}

         \begin{figure}[p]
         \includegraphics[width=6.5in]{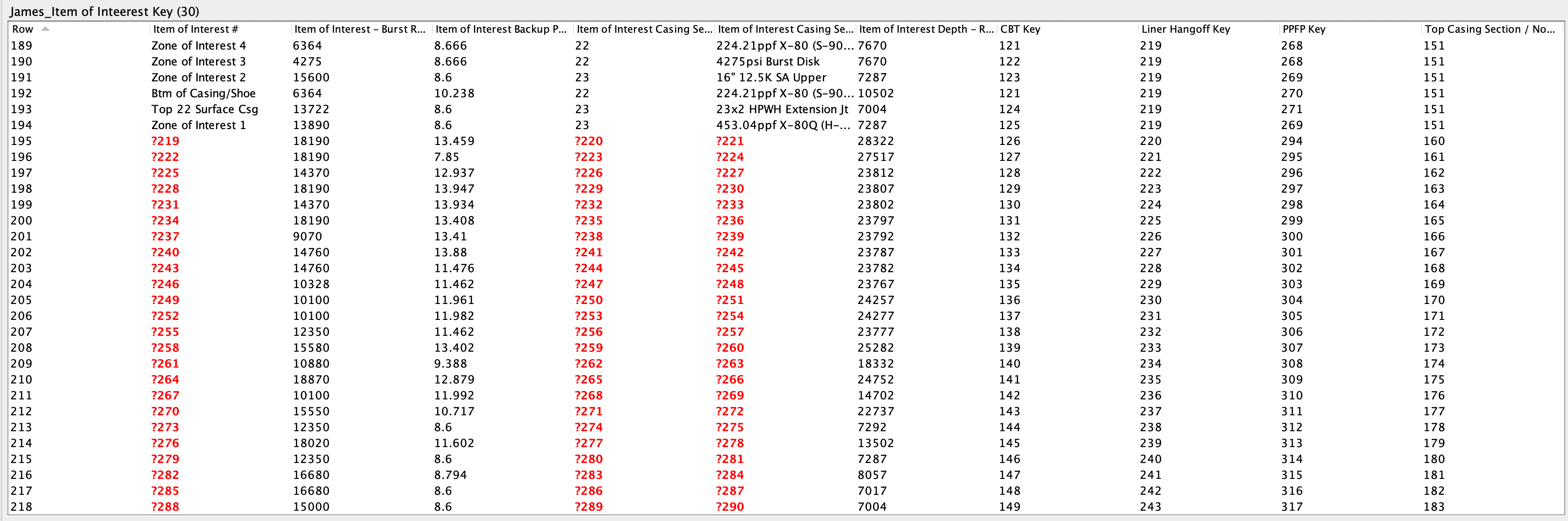}
         \includegraphics[width=6.5in]{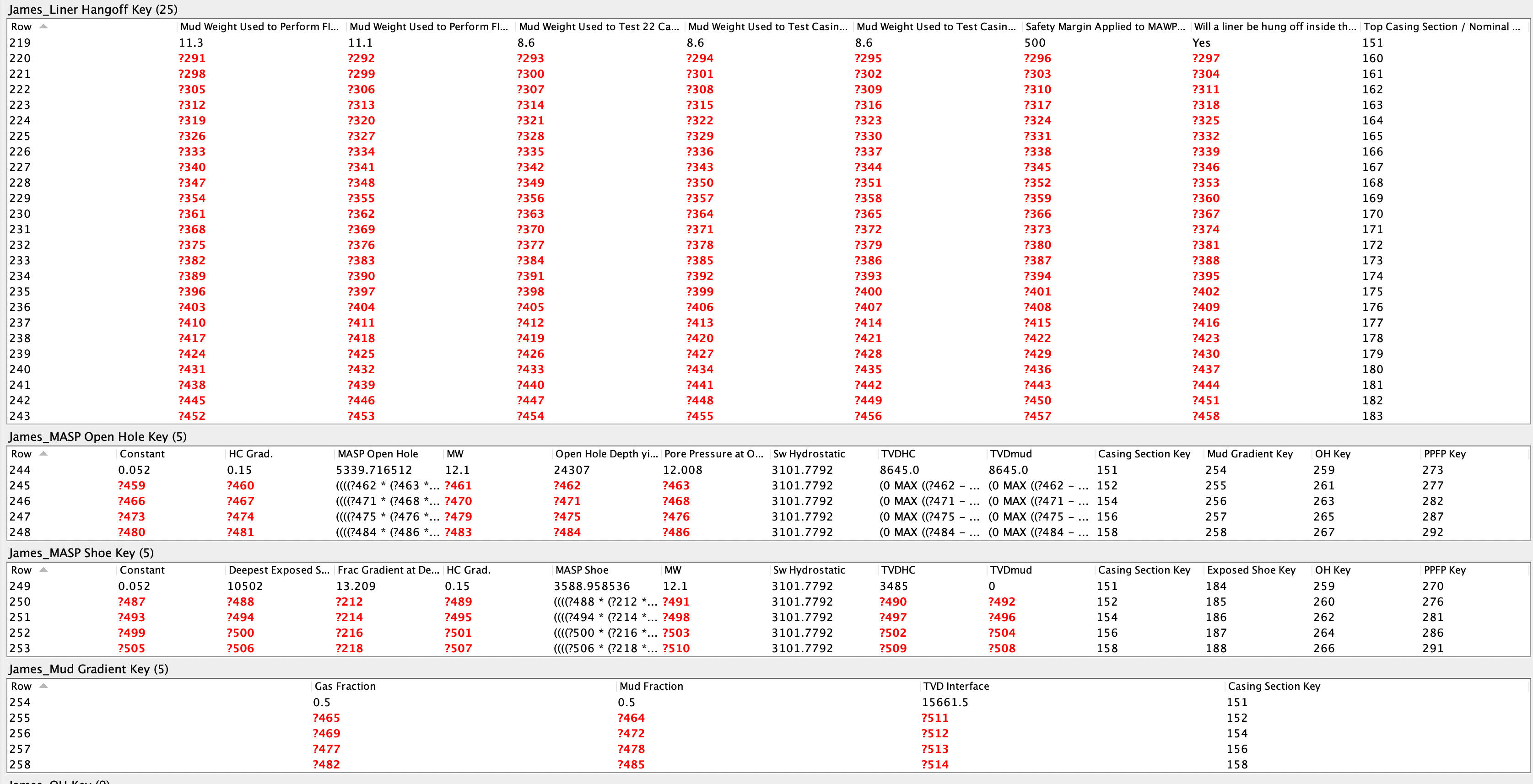}

\includegraphics[width=6.5in]{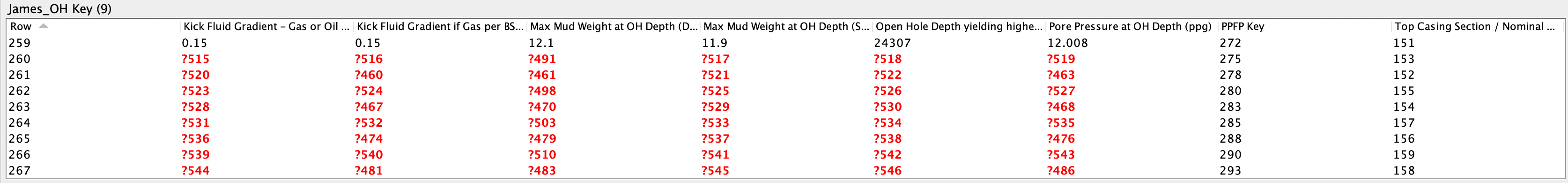}

        \caption{Full Integrated Olog, 3 of 4}
        \label{fig:intD}
    \end{figure}
    
     \begin{figure}[p]
     \vspace*{-.5in}
\includegraphics[width=6.5in]{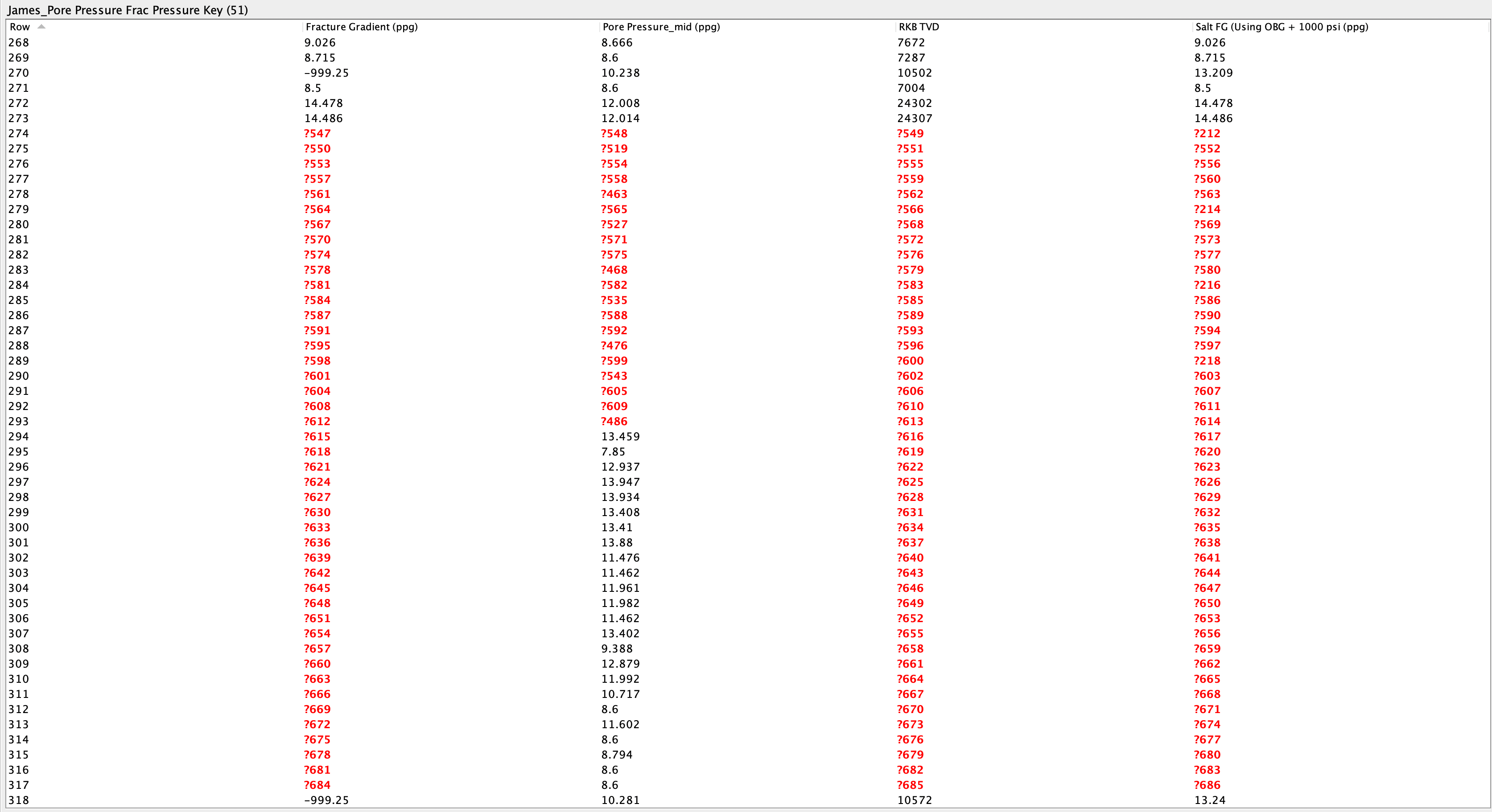}
\includegraphics[width=6.5in]{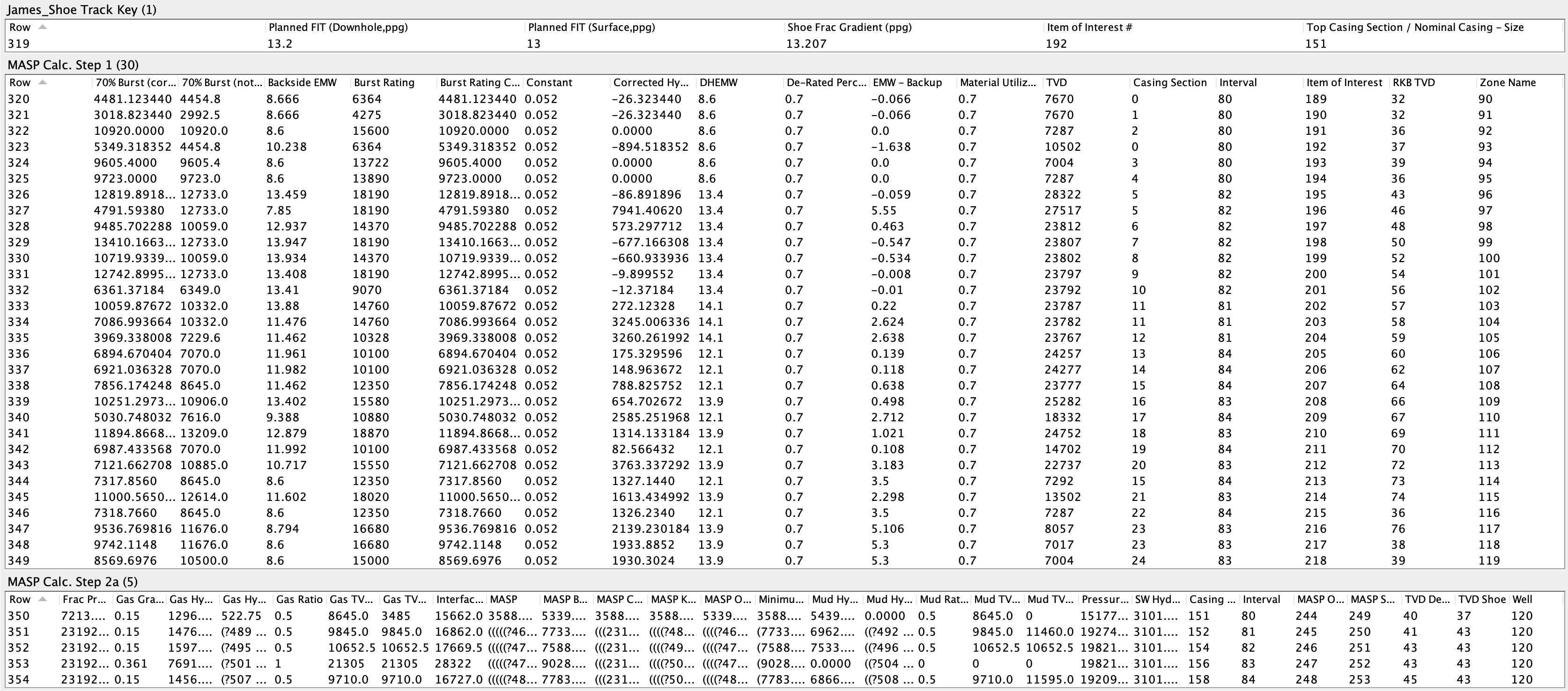}

        \caption{Full Integrated Olog, 4 of 4}
        \label{fig:intE}
    \end{figure}

\newpage 

\newpage

\begin{footnotesize}
\vspace*{-.5in}
\tableofcontents
\listoffigures
\end{footnotesize}

\end{document}